\documentclass[letter,11pt]{article}
\pdfoutput=1 % if your are submitting a pdflatex (i.e. if you have
\usepackage{jheppub} % for details on the use of the package, please
\usepackage[T1]{fontenc} % if needed

\usepackage{url,comment}
\usepackage{times}
\usepackage{latexsym}
\usepackage{graphicx, graphics, hyperref, amsmath, amssymb, slashed, xcolor, bbm,amsthm, array,textcomp,gensymb}
 \usepackage{subfigure}
\newcolumntype{P}[1]{>{\centering\arraybackslash}p{#1}}
\newcolumntype{M}[1]{>{\centering\arraybackslash}m{#1}}
\usepackage{rotating}
\usepackage{afterpage}

\usepackage{enumitem}

\newcommand{\nc}{\newcommand}
\nc{\beq}{\begin{equation}}
\nc{\eeq}{\end{equation}}
\nc{\barray}{\begin{eqnarray}}
\nc{\earray}{\end{eqnarray}}
\nc{\barrayn}{\begin{eqnarray*}}
\nc{\earrayn}{\end{eqnarray*}}
\nc{\bcenter}{\begin{center}}
\nc{\ecenter}{\end{center}}
\nc{\mc}{\mathcal}
\nc{\er}[1]{(\ref{eq:#1})}
\nc{\onehalf}{\frac{1}{2}} 
\nc{\partialbar}{\bar{\partial}}
\nc{\psit}{\widetilde{\psi}}
\nc{\Tr}{\mbox{Tr}}
\nc{\hc}{\mbox{H.c.}}
\nc{\ev}{\;\mathrm{eV}}
\nc{\mev}{\;\mathrm{MeV}}
\nc{\gev}{\;\mathrm{GeV}}
\nc{\tev}{\;\mathrm{TeV}}

\def\chii0{\chi_i^0}
\def\chij0{\chi_j^0}

\newcommand{\gsim}{\lower.7ex\hbox{$\;\stackrel{\textstyle>}{\sim}\;$}}% This is just \gtrsim
\newcommand{\lsim}{\lower.7ex\hbox{$\;\stackrel{\textstyle<}{\sim}\;$}}%This is just \lesssim
\nc{\ttbar}{t\bar t}

\def\iab{{\ \rm ab}^{-1}}

\newcommand{\fref}[1]{Fig.~\ref{f.#1}}
\newcommand{\eref}[1]{Eq.~(\ref{e.#1})}

\newcommand{\aref}[1]{Appendix~\ref{a.#1}}
\newcommand{\sref}[1]{Section~\ref{s.#1}}
\newcommand{\ssref}[1]{Section~\ref{ss.#1}}
\newcommand{\sssref}[1]{Section~\ref{sss.#1}}
\newcommand{\cref}[1]{Chapter~\ref{c.#1}}

\graphicspath{{figs/}}

\title{\boldmath 
Thermal Resummation and Phase Transitions
}
\preprint{YITP-2016-48}
\author[a]{David Curtin}
\author[b]{Patrick Meade}
\author[b]{Harikrishnan Ramani}

\affiliation[a]{Maryland Center for Fundamental Physics,
University of Maryland, College Park, MD 20742, USA}
\affiliation[b]{C. N. Yang Institute for Theoretical Physics, SUNY Stony Brook, Stony Brook, NY 11794, USA}

\emailAdd{dcurtin1@umd.edu}
\emailAdd{patrick.meade@stonybrook.edu}
\emailAdd{harikrishnan.ramani@stonybrook.edu}

\abstract{
The consequences of phase transitions in the early universe are becoming testable in a variety of manners, from colliders physics to gravitational wave astronomy.   In particular one phase transition we know of, the Electroweak Phase Transition (EWPT), could potentially be first order in BSM scenarios and testable in the near future.  If confirmed this could provide a mechanism for Baryogenesis, which is one of the most important outstanding questions in physics.  To reliably make predictions it is necessary to have full control of the finite temperature scalar potentials.  However, as we show the standard methods used in BSM physics to improve  phase transition calculations, resumming hard thermal loops, introduces significant errors into the scalar potential.  In addition, the standard methods make it impossible to match theories to an EFT description reliably.  In this paper we define a thermal resummation procedure based on Partial Dressing (PD) for general BSM calculations of phase transitions beyond the high-temperature approximation.  Additionally, we introduce the modified Optimized Partial Dressing (OPD) procedure, which is numerically nearly as efficient as old incorrect methods, while yielding identical results to the full PD calculation. This can be easily applied to future BSM studies of phase transitions in the early universe.  As an example, we show that in unmixed singlet scalar extensions of the SM, the (O)PD calculations make new phenomenological predictions compared to previous analyses. An important future application is the study of EFTs at finite temperature.
}

\begin{document} 
\maketitle
\flushbottom

%%%%%%%%%%%%%%%%%%%%%%%%%%%%
%%%%%%%%%%%%%%%%%%%%%%%%%%%%
%%%%%%%%%%%%%%%%%%%%%%%%%%%%
\section{Introduction}
\label{s.intro}
%%%%%%%%%%%%%%%%%%%%%%%%%%%%
%%%%%%%%%%%%%%%%%%%%%%%%%%%%
%%%%%%%%%%%%%%%%%%%%%%%%%%%%

Thermal phase transitions are ubiquitous phenomena in nature, but in fundamental physics they are difficult to study, and very few are known.  In the SM there are phase transitions associated with QCD and the EW symmetry.  The QCD phase transition can be studied directly in heavy ion collisions, with rapid progress over the last decade~\cite{Akiba:2015jwa}.  However, the EW phase transition (EWPT) is far out of reach of direct testability.   Going beyond the Standard Model (BSM), the nature of the EW phase transition could change, and there could be additional phase transitions unrelated to the EWPT but still screened from us by the CMB.  
Nevertheless, even without direct measurements of the EWPT in the near future these phenomena can be indirectly studied, with profound consequences  for our understanding of the early universe.
 The EWPT and other phase transitions can have correlated signals detectable  at current and future colliders, and in the burgeoning field of gravitational wave astronomy.  
 Therefore, it is important to have as much control of the underlying Finite-Temperature Quantum Field Theory (FTQFT) calculations as possible, so that potential signals are reliably understood and predicted.  This is the aim of this paper, and we will introduce new methods in FTQFT to capture the effects of BSM physics on phase transitions. While our results will be general, we  single out the EWPT for special study given its possible deep connection to another fundamental question in particle physics.  

One of the most profound mysteries in particle physics is our mere existence, and that of all baryons in the universe.  A dynamical explanation for our universe containing an excess of matter over antimatter requires BSM physics. At some time in the history of the primordial plasma, after reheating but before Big Bang Nucleosynthesis ($T \sim \gev$), a mechanism of \emph{baryogenesis} has to create the observed baryon asymmetry~\cite{Ade:2015xua, Kolb:1990vq} of
\begin{equation}
\eta = \frac{n_B - n_{\overline B}}{n_\gamma} \sim 10^{-9} \ .
\end{equation}
This requires the three Sakharov conditions \cite{Sakharov:1967dj} to be satisfied: baryon number ($B$) violation, $CP$ violation, and a sufficiently sharp departure from thermal equilibrium. 

\emph{Electroweak Baryogenesis} \cite{
Kuzmin:1985mm, Klinkhamer:1984di, Shaposhnikov:1986jp, Shaposhnikov:1987tw, Arnold:1987mh, Arnold:1987zg, Khlebnikov:1988sr} is a very appealing possibility, since all involved processes must occur near the weak scale, making it in principle testable. (See \cite{Cline:2006ts, Trodden:1998ym, Riotto:1998bt, Riotto:1999yt, Quiros:1999jp, Morrissey:2012db} for reviews.)
In the SM, high temperature effects stabilize the Higgs field at the origin, restoring electroweak symmetry~\cite{Dolan:1973qd, Weinberg:1974hy}. In this high-temperature unbroken phase, the SM in fact contains a $(B+L)$-violating process in the form of nonperturbative sphaleron transitions, which can convert a chiral asymmetry into a baryon asymmetry. The EWPT from the unbroken to the broken phase at $T \sim 100 \gev$ provides, in principle, a departure from thermal equilibrium. In the presence of sufficient $CP$-violation in the plasma, a baryon excess can be generated

EWBG cannot function within the SM alone. There is insufficient $CP$-violation (see for example ~\cite{Riotto:1998bt}), and the EWPT is not first order for $m_h \gtrsim 70 \gev$ \cite{Bochkarev:1987wf, Kajantie:1995kf}. Additional BSM physics is required to generate a strong phase transition (PT) and supply additional $CP$-violating interactions in the plasma. 

Many theories have been proposed to fulfill these requirements of EWBG, 
 including extensions of the scalar sector with additional singlets
\cite{Profumo:2007wc, Ashoorioon:2009nf, Damgaard:2013kva,  Barger:2007im, Espinosa:2011ax, Noble:2007kk, Cline:2012hg, Cline:2013gha, Alanne:2014bra, Espinosa:2007qk, Profumo:2014opa, Fuyuto:2014yia,  Fairbairn:2013uta, Jiang:2015cwa} (which can be embedded in supersymmetric models \cite{Pietroni:1992in, Davies:1996qn, Huber:2006wf, Menon:2004wv, Huber:2006ma, Huang:2014ifa, Kozaczuk:2014kva}),
two-Higgs doublet models \cite{Dorsch:2013wja, Dorsch:2014qja}, triplet extensions \cite{Patel:2012pi, Blinov:2015sna, Inoue:2015pza}, and the well-known Light-Stop Scenario in the MSSM \cite{Carena:1996wj, Laine:1998qk, Espinosa:1996qw, Delepine:1996vn, Carena:1997ki, Huber:2001xf, Carena:2002ss, Lee:2004we, Carena:2008rt, Carena:2008vj, Cirigliano:2009yd} which is now excluded \cite{Curtin:2012aa, Cohen:2012zza, Katz:2014bha, Katz:2015uja}.
To determine whether a particular, complete model can successfully account for the observed baryon asymmetry, the temperature-dependent Higgs potential and the resulting nature of the phase transition have to be carefully calculated to determine the sphaleron energy as well as the bubble nucleation rate and profile. 
This information serves as an input to solve a set of plasma transport equations, which determine the generated baryon asymmetry of the universe (BAU). 
The full calculation is very intricate, with many unresolved theoretical challenges (see e.g. \cite{Huber:2001xf, Cline:2000kb, Moore:2000wx, John:2000zq, John:2000is, Megevand:2009gh, Moreno:1998bq, Riotto:1998zb, Lee:2004we, Cirigliano:2006wh, Chung:2008aya, Carena:2002ss, Li:2008ez, Cline:1997vk, Cline:2000nw, Konstandin:2004gy, Konstandin:2003dx, Kozaczuk:2012xv}).

The sectors of a theory which generate the strong phase transition, and generate baryon number via $CP$-violating interactions in the plasma, do not have to be connected (though they can be). Since one of the most appealing features of EWBG is its testability, it makes sense to consider these two conditions and their signatures separately. The ultimate aim is a model-independent understanding of the collider, low-energy, and cosmological signatures predicted by all the various incarnations of EWBG.

We will focus on the strong electroweak phase transition. If it is first order, there is a critical temperature $T_c$ where the Higgs potential has two degenerate minima $h = 0$ and $v_c$, separated by an energy barrier. As the temperature decreases, the minimum away from the origin becomes the true vacuum, the Higgs field tunnels to the broken phase, and bubbles of true vacuum expand to fill the universe. A necessary condition for avoiding baryon washout is that $v_c$ is sufficiently large to suppress sphalerons. Specifically, a BSM theory which realizes EWBG has to satisfy
\begin{equation}
\frac{v_c}{T_c} > 0.6 - 1.6 \ .
\end{equation}
In most cases we will adopt the lower value of 0.6 as our cutoff \cite{Patel:2011th} to be as inclusive as possible (though it is sometimes instructive to examine the parameter space that survives the more standard $v_c/T_c > 1$ criterion.) This is a useful way of checking whether a given BSM scenario is a viable candidate for EWBG, as well as determining the correlated signatures we could measure today.

Computing this ratio seems like a straightforward exercise, and a standard recipe has been adopted in the literature for computing the EWPT in BSM models (see e.g. \cite{Quiros:1999jp} for a review). This involves constructing the one-loop effective Higgs potential at finite temperature by using a well-known generalization of the standard Coleman-Weinberg potential; possibly including a selection of the most important higher-loop effects and/or RG-improvements;  and resumming an important set of contributions called hard thermal loops.

We carefully review this calculation in \sref{review}. Our focus is the resummation of hard thermal loops. The  standard procedure, which we call \emph{Truncated Full Dressing} (TFD), involves a very simple computation of thermal masses $\Pi_i \sim T^2$ for particles $i$ in the plasma, to leading order in the high-temperature approximation, and inserting them back into the effective potential \cite{Gross:1980br, Parwani:1991gq, Arnold:1992rz}.
%\DC{original daisy resum refs}. 

Various extensions of this simple recipe, to include higher-order corrections in temperature or coupling, have been explored roughly twenty years ago in the context of $\phi^4$ theories \cite{Espinosa:1992gq, Espinosa:1992kf, Quiros:1992ez, Boyd:1993tz, Dine:1992wr, Boyd:1992xn}. However, possibly because the consensus on the (most) correct generalization seemed unclear, and the involved calculations seemed  onerous to perform for every BSM theory, these improvements have not found wide application in  the study of the EWPT in general BSM scenarios. 

We revisit these issues in a modern light, with a focus on the study of general BSM effects which can induce a strong EWPT.  A simple and easily implementable extension of the TFD calculation is urgently  required for two reasons: to correctly determine the phenomenology of EWBG, and to understand Effective Field Theory (EFT) at finite temperature. 

 Since the high-temperature expansion of $\Pi_i$ is truncated at the leading term, it is only accurate to $\mathcal{O}(m/T)$. This can easily be $\sim 40\%$ at $T = T_c$, and vary with the Higgs field since its VEV determines particle masses.  
While this does not directly translate to a corresponding error on the full effective potential, an important class of (particularly testable) EWBG theories generates a strong EWPT via a partial cancellation between $\Pi_i$ and tree-level parameters. In this case, accurate determination of the thermal masses, and their $h$-dependence, is clearly necessary to have confidence in the results of the phase transition computation, and hence the observables correlated with EWBG.

Effective Field Theories (EFTs) are a powerful tool to parameterize general new physics effects at zero temperature as a set of non-renormalizable operators involving SM fields.
To understand the signatures of EWBG in a model-independent fashion, one would like to extend such an EFT analysis to finite temperature. Early attempts like \cite{Grojean:2004xa, Bodeker:2004ws, Delaunay:2007wb} suggested that a $|H|^6$ operator could induce a strong EWPT in correlation with sizable deviations in the cubic Higgs self-coupling, which could be detected with the next generation of future lepton \cite{Asner:2013psa, Tian:2013yda} and 100 TeV \cite{Tang:2015qga, Contino:2016spe} colliders, or even the HL-LHC \cite{ATLAS:lambda3, Baglio:2012np, Goertz:2013kp, Barger:2013jfa,Yao:2013ika}. 
Unfortunately, EFTs at finite Temperature are very poorly understood. For example, the effects of a particle with a mass of 300 GeV are quite well described in an EFT framework for collider experiments with $\sqrt{s} = 100 \gev$, but it seems doubtful that this is the case for temperatures of $T = 100 \gev$, since thermal fluctuations can excite modes somewhat heavier than $T$. Without understanding these effects in detail, we cannot know the EFT's radius of convergence in field space or temperature, and hence know whether its predictions regarding the EWPT can be trusted. 
Since the assumptions of the high-temperature approximation ($T \gg m$) for $\Pi_i$ in TFD are fundamentally incompatible with the assumptions of an finite-temperature EFT analysis ($T \ll m$), careful study of these decoupling effects during a phase transition, and EFT matching at finite temperature, requires a more complete treatment of thermal masses.

In this work, we develop a consistent, easily implementable procedure for the numerical computation and resummation of thermal masses in general BSM theories, beyond leading order in temperature and coupling. 

We examine two competing approaches which were proposed in the context of $\phi^4$ theories: \emph{Full Dressing} (FD) \cite{Espinosa:1992gq, Espinosa:1992kf, Quiros:1992ez} and \emph{Partial Dressing} (PD) \cite{Boyd:1993tz}. 
We verified the claims of \cite{Boyd:1993tz} that PD avoids the problem of miscounting diagrams beyond one-loop order~\cite{Dine:1992wr, Boyd:1992xn}, and that it generalizes beyond $\mathcal{O}(T^2)$. 
We therefore focus on PD. We review its formal underpinnings in \sref{formal} and outline how to generalize it to BSM theories, in general without relying on the high-temperature approximation.

Applying the PD procedure beyond the high-temperature approximation requires numerically solving a type of finite-temperature gap equation. We outline the implementation of this calculation in \sref{calculation} in the context of a specific BSM benchmark model. Computing the strength of the EWPT with PD is extremely numerically intensive, necessitating the use of a custom-built \texttt{c++} code. This allows us to study the importance of resummed finite-temperature effects for the phase transition, but is impractical for future BSM studies. 
We show it is possible to modify PD by extending the gap equation and implementing certain approximations. This greatly increases numerical reliability, while reducing CPU cost by several orders of magnitude. We call this updated resummation procedure \emph{Optimized Partial Dressing} (OPD) and show it is equivalent to PD for BSM studies of the EWPT. 

OPD is only slightly more CPU-intensive than the standard TFD calculation used in most studies of the EWPT to date, and very easy to implement in Mathematica. We hope that this calculation, which is explained in \sref{calculation} and summarized in the form of an instruction manual in \aref{instructions}, will be useful in the future study of the EWPT for BSM theories.

The BSM model we use to develop and evaluate the PD and OPD resummation schemes is the SM with $N_S$ added singlets transforming under an unbroken $O(N_S$) symmetry in our vacuum (or $\mathbb{Z}_2$ if $N_S = 1$) and coupling to the SM via a quartic Higgs portal $\lambda_{HS} |H|^2 S_i^2$ without Higgs mixing. This benchmark model serves as a useful ``worst-case scenario'' for the collider phenomenology of EWBG, since it can produce a strong EWPT in a variety of ways which are representative of more complete theories, while generating the minimal set of collider signatures consistent with EWBG. The authors of \cite{Curtin:2014jma} studied this scenario with $N_S = 1$ using the TFD calculation, making progress towards a ``phenomenological no-lose theorem for EWBG'' by showing that the future 100 TeV and lepton colliders could probe its EWBG-compatible parameter space completely. We update and generalize this phenomenological analysis for $N_S \geq 1$ in the PD scheme. 
As shown in \sref{physical}, the ``no-lose theorem'' is strengthened, with EWPTs caused by larger numbers of scalars being easier to detect at colliders. 

Phenomenologically, the main lessons of the updated PD calculation are that the detailed correlations between a strong EWPT and collider observables can be significantly shifted, especially in more complete theories of EWBG than our SM + $N_S \times S$ benchmark model. Furthermore, two-step transitions are more prevalent than suggested by earlier TFD calculations. This raises the exciting prospect of discovering the traces of a strong two-step transition with gravitational wave observations \cite{Kamionkowski:1993fg, Grojean:2006bp}. Finally, unlike (O)PD, the TFD calculation overestimates the reliability of the finite-temperature EWPT calculation, underlining the importance of tracking error terms when computing the strength of the PT. 

This paper is structured as follows. The standard TFD calculation of the EWPT is pedagogically reviewed in \sref{review}. \sref{formal} lays the formal groundwork of the PD scheme, while the implementation of the full calculation and its extension to the OPD scheme is described in \sref{calculation}. The differences in physical predictions between the standard TFD and the new (O)PD calculation are explored in \sref{physical}. We conclude in \sref{conclusions}, and provide an instruction manual for easy implementation of the OPD calculation for the EWPT in \aref{instructions}.

%%%%%%%%%%%%%%%%%%%%%%%%%%%%
%%%%%%%%%%%%%%%%%%%%%%%%%%%%
%%%%%%%%%%%%%%%%%%%%%%%%%%%%
\section{Review: Calculating the Electroweak Phase Transition}
\label{s.review}
%%%%%%%%%%%%%%%%%%%%%%%%%%%%
%%%%%%%%%%%%%%%%%%%%%%%%%%%%
%%%%%%%%%%%%%%%%%%%%%%%%%%%%

We now review the standard computation of the finite-temperature Higgs potential in BSM theories (see e.g.~\cite{Quiros:1999jp}). We call the leading-order thermal mass resummation~\cite{Gross:1980br, Parwani:1991gq, Arnold:1992rz} TFD, to contrast with the PD procedure which we review and develop further in Sections \ref{s.formal} and \ref{s.calculation}. This will make plain some important shortcomings of TFD. 

As a BSM benchmark, we consider the SM with $N_S$ added real SM-singlet scalar fields $S_i$ obeying an $O(N_S)$ symmetry (or $\mathbb{Z}_2$ if $N_S = 1$). We are also interested in regions of parameter space where this symmetry is unbroken in the zero-temperature vacuum of our universe today (i.e. $\langle S_i \rangle = 0$). This forbids Higgs-Singlet mixing, which significantly simplifies several formal aspects of thermal mass resummation. Unmixed singlet extensions also represent a useful ``phenomenological nightmare scenario'' for EWBG \cite{Curtin:2014jma} with minimal experimental signatures. We show in \sref{physical} that this model can nonetheless be completely probed by the next generation of colliders.

\subsection{Tree-level Potential}

 The tree-level scalar potential is
\begin{equation}
\label{e.V0full}
V_0 = - \mu^2 |H|^2 + \lambda |H|^4  + \frac{1}{2} \mu_S^2 (S_i S_i) + \frac{1}{4} \lambda_S (S_i S_i)^2 + \lambda_{HS} |H|^2 (S_i S_i).
\end{equation}
We focus on the real $h$ component of the SM Higgs doublet $H = (G^+ , (h + iG^0 )/\sqrt{ 2})$ which acquires a VEV during EWSB. Without loss of generality, we also assume that any excursion in S-field-space occurs along the $S_0$ direction. Therefore, the relevant part of the tree-level potential is
\begin{equation}
\label{e.V0hS0}
V_0 = - \frac{1}{2} \mu^2 h^2 + \frac{1}{4} \lambda h^4  + \frac{1}{2} \mu_S^2 S_0^2  + \frac{1}{4} \lambda_S S_0^4 + \frac{1}{2} \lambda_{HS} h^2 S_0^2 \ .
\end{equation}
(Of course, \eref{V0full} determines the form of the scalar masses $m^2_{G^+}(h,S_0), m^2_{S_i}(h,S_0), \ldots$ which determine the form of one-loop contributions as outlined below.) 
Our aim is to obtain the effective potential $V_\mathrm{eff}(h, S, T)$ at one-loop order.

\subsection{Coleman Weinberg Potential}
At zero-temperature, the one-loop effective potential can be written as 
\begin{equation}
\label{e.Vcw}
V_\mathrm{eff}^{T=0}(h,S) = V_0 + \sum_i V^i_\mathrm{CW}(m_i^2(h,S))
\end{equation}
The Coleman-Weinberg potential is the zero-momentum piece of the zero-temperature effective action, and is a sum of 1PI one-loop diagrams with arbitrary numbers of external $h$ and $S_0$ fields and particles $i = \{t, W, Z, h, G^\pm, G^0, S_0, S_k, \ldots\}$ running in the loop (where $k > 0$). Note that we are working in Landau gauge to avoid ghost-compensating terms, which requires including the Goldstone contributions separately, in addition to the massive $W$, $Z$ bosons. (We discuss issues of gauge invariance in \ssref{problems}.)

The dependence of the $i^\mathrm{th}$ particle tree-level mass $m_i^2 = m_i^2(h,S)$ on the VEVs of $h$ and $S$ determines $V_\mathrm{CW}$. Summing over all contributions gives~\cite{Coleman:1973jx}:
\begin{equation}
V_{CW}^i=\frac{1}{2} \int \frac{d^4k}{(2\pi)^4} \log[k^2_E+m_i^2(h,S)]
\end{equation}
where e.g. $m_h(h,S_0) = - \mu^2 + 3 \lambda h^2 + \lambda_{HS} S_0^2$ and $k_E$ is the euclidian momentum of particle $i$ in the loop. We adopt the dimensional regularization scheme and $\overline{\mathrm{MS}}$ renormalization scheme, with the usual $\epsilon=2-\frac{D}{2}$. This makes one-loop matching more onerous than the on-shell renormalization scheme, but allows for the potential to be RG-improved more easily. The result is
\begin{equation}
V^i_{CW}=\frac{m_i^4(h,S)}{64 \pi^2}\left(-\epsilon-\gamma_E+\log 4\pi+ \log \frac{m_i^2(h,S)}{\mu^2_R}\right)
\end{equation}
(Hereafter we drop the explicit $h,S$ dependence of the masses for brevity.) $\mu_R$ is the renormalization scale, and variation of physical observables after matching with different values of $\mu_R$ is a common way of assessing the uncertainty of our results due to the finite one-loop perturbative expansion. Adding counterterms and removing divergences yields the familiar expression
\begin{equation}
V^i_{CW}=(-1)^F g_i \frac{m_i^4}{64 \pi^2} \left(\log\left[\frac{m_i^2}{\mu_R^2}\right]-c_i\right)
\label{e.CW0} ,
\end{equation}
where $F=1 (0)$ for fermions  (bosons), $c_i=\frac{3}{2} (\frac{5}{2}$) for scalars/fermions (vectors), and $g_i$ is the number of degrees of freedom associated with the particle $i$.

\subsection{Finite Temperature}
\label{ss.finitetemperature}

Finite-temperature quantum field theory (FTQFT) enables the computation of observables, like scalar field  vacuum expectation values, in the background of a thermal bath. The corresponding Greens functions can be computed by compactifying time along the imaginary direction, for details see e.g.~\cite{Quiros:1999jp}. 
To get an intuitive idea of finite-temperature effects on the one-loop effective potential, it is useful to consider integrals of the form
\begin{equation}
\int \frac{d k_0}{2\pi} f(k_0)
\end{equation}
where $k_0$ is the time-like component of the loop momentum. This can be evaluated in FTQFT as
\begin{equation}
\label{e.FTQFT1}
\int \frac{d k_0}{2\pi} f(k_0) \rightarrow T \sum_{n = -\infty}^{\infty} f(k_0 = i \omega_n) \ ,
\end{equation}
where $\omega_n=2 n  \pi T$ and $(2n + 1)  \pi T$ are the Matsubara frequencies for bosons and fermions, respectively. \eref{FTQFT1} can be written in the instructive form:
\begin{equation}
\int \frac{d k_0}{2\pi} f(k_0) \rightarrow \int^{i\infty}_{-i\infty} \frac{dz}{4 \pi i} \left[f(z) + f(-z)\right]  \ + \ \eta \int_C \frac{dz}{2 \pi i} n(z) \left[f(z) + f(-z)\right] \ ,
\end{equation}
where $\eta = \pm 1$ for bosons/fermions and $n(z) = (e^{z/T} - \eta)^{-1}$ are the standard Fermi-Dirac/Bose-Einstein distribution functions, for a particular choice of contour $C$. The first term, which is $n(z)$ independent, is simply the usual zero-temperature loop integral, while the second term is the new contribution from thermal loops in the plasma. This makes effects like thermal decoupling very apparent -- if the particle mass is much larger than the temperature, its contribution to the second loop integral will vanish as $n(z) \to 0$.

Applying this formalism to the one-loop effective potential at finite temperature generalizes \eref{Vcw}:
\begin{equation}
\label{e.VefffiniteT}
V_\mathrm{eff}^{T>0}(h,S,T) = V_0 + 
 \sum_i \left[ 
V^i_\mathrm{CW}(m_i^2(h,S)) + V^i_\mathrm{th}(m_i^2(h,S),T)
\right]
\ ,
\end{equation}
where the second term is the usual Coleman-Weinberg potential, and the third term is the one-loop thermal potential
\begin{equation}
V^i_\mathrm{th}(m_i^2(h,S),T)
= (-1)^F g_i  \frac{T^4}{2 \pi^2} J_{\text{B/F}}\left(\frac{m_i^2(h,S)}{T^2}\right)
\label{e.Vth}
\end{equation}
with thermal functions
\begin{equation}
J_\text{B/F}(y^2) = \int_0^\infty dx \ x^2 \ \log\left[1 \mp \exp (- \sqrt{x^2 + y^2}) \right] 
\label{e.JBF}
\end{equation}
which vanish as $T \to 0$. Note that $y^2$ can be negative. 
The thermal functions have very useful closed forms in the high-temperature limit,
\begin{eqnarray}
\nonumber
J_B(y^2) &\approx& 
J_B^{\mathrm{high}-T}(y^2) = 
-\frac{\pi^4}{45} + \frac{\pi^2}{12} y^2  - \frac{\pi}{6} y^3 - \frac{1}{32} y^4 \log\left( \frac{y^2}{a_b}\right)
\\
\label{e.JBFhighT}
J_F(y^2) &\approx&
J_F^{\mathrm{high}-T}(y^2) = 
 \frac{7 \pi^4}{360} - \frac{\pi^2}{24} y^2 - \frac{1}{32} y^4 \log\left(\frac{y^2}{a_f}\right) 
 \ \ \ \ \ \ \ \ \ \ \ \ \ \  \mathrm{for} \ |y^2| \ll 1
\ ,
\end{eqnarray}
where $a_b = \pi^2 \exp(3/2 - 2 \gamma_E)$ and $a_f = 16 \pi^2 \exp(3/2 - 2 \gamma_E)$. 
This high-$T$ expansion includes more terms, but they do not significantly increase the radius of convergence. With the log term included, this approximation for both the potential and its derivatives is accurate to better than $\sim 10\%$ even for $m \sim (1 - 3) \times T$ (depending on the function and order of derivative), but breaks down completely beyond that. 
The low-temperature limit ($|y^2| \gg 1$) also has a useful expansion in terms of modified Bessel functions of the second kind:
\begin{eqnarray}
\nonumber
J_B(y^2) &=& \tilde J_B^{(m)}(y^2) = - \sum_{n=1}^m \frac{1}{n^2} y^2 K_2(y \ n)
\\
\label{e.JBFlowT}
J_F(y^2) &=&
 \tilde J_F^{(m)}(y^2) = 
 - \sum_{n=1}^m \frac{(-1)^n}{n^2} y^2 K_2(y \ n)
 \ \ \ \ \ \ \ \ \ \ \ \ \ \ \ \ \ \ \ \ \ \ \ \ \  \mathrm{for} \ |y^2| \gg 1
\ .
\end{eqnarray}
This expansion can be truncated at a few terms, $m = 2$ or 3, and yield very good accuracy, but convergence for smaller $|y^2|$ is improved by including more terms.

For negative $m_i^2$, the effective potential (both zero- and finite-temperature) includes imaginary contributions, which were discussed in~\cite{Delaunay:2007wb}. These are related to decay widths of modes expanded around unstable regions of field space, and do not affect the computation of the phase transition. Therefore, we always analyze only the real part of the effective potential.

\subsection{Resummation of the Thermal Mass: Truncated Full Dressing (TFD)}
\label{ss.standardresummation}

The effective potential defined in \eref{VefffiniteT} can be evaluated at different temperatures to find $v_c$ and $T_c$. 
The high-temperature expansion of $J_B$ already reveals how one particular BSM effect could induce a strong EWPT. If a light boson is added to the plasma with $m_i^2 \sim h^2$, then the $-y^3$ term in $J_B$ will generate a negative  cubic term $-h^3$  in the effective potential, which  can generate an energy barrier between two degenerate vacua. 
However, the calculation of the effective finite-temperature potential is still incomplete. There is a very well-known problem which must be addressed in order to obtain a trustworthy calculation \cite{Gross:1980br, Parwani:1991gq, Arnold:1992rz}. 

This can be anticipated from the fact that a symmetry, which is broken at zero temperature, is restored at high-temperature. Thermal loop effects overpower a temperature-independent tree-level potential. 
This signals a breakdown of fixed-order perturbation theory, which arises because in FTQFT a massive scalar theory has not one but two scales: $\mu$ and $T$. Large ratios of $T/\mu$ have to be resummed.

To begin discussing thermal mass resummation, let us first consider not our BSM benchmark model but a much simpler $\phi^4$ theory with quartic coupling $\lambda$ and $N$ scalar fields which obey a global $O(N)$ symmetry:
\begin{equation}
V_0 = - \frac{1}{2} \mu^2 \phi_i \phi_i + \frac{1}{4} \lambda (\phi_i \phi_i)^2 \ .
\end{equation}
(In fact, if we were to ignore fermions and gauge bosons,  and set $\lambda = \lambda_{HS} = \lambda_S$, our BSM benchmark model would reduce to this case with $N = N_S + 4$.) 

\begin{figure}
\centering
\begin{tabular}{cccc}
\includegraphics[width=0.15\textwidth]{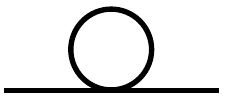}
&
\includegraphics[width=0.15\textwidth]{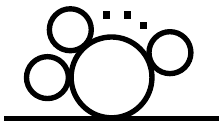} 
&
\includegraphics[width=0.15\textwidth]{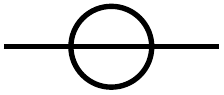} 
&
\includegraphics[width=0.15\textwidth]{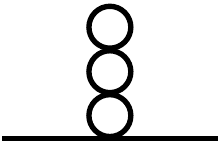} 
\\
(a) & (b) & (c) & (d)
\end{tabular}
\caption{
Various scalar mass contributions in $\phi^4$ theory:
(a)~one-loop mass correction, which is quadratically divergent at zero temperature,
(b)~higher-loop daisy contributions which are leading order in $T$ and $N$ at high temperature, 
(c)~the two-loop ``lollipop'' contribution which is subleading in $T$ and $N$ to the two-loop daisy,
(d)~the three-loop superdaisy contribution, which is subleading in $T$ but of equal order in $N$ to the three-loop daisy.
%one-loop thermal mass corresponding to the quadratically divergent  Ring diagrams to all order captured by resummation. Daisy diagrams
}
\label{f.daisyresum}
\end{figure}

In FTQFT, the leading order high temperature behavior for diagrams with external scalar lines is directly related to the 0-temperature superficial degree of divergence $d$. Diagrams which have $d>0$ have a $T^d$ high temperature behavior. For $d \le 0$, there is a linear $T$ dependence. Appropriate factors of the coupling $\lambda$, $N$ and the tree level mass parameter $\mu$ can be added from vertex counting and dimensional grounds. Therefore, the one-loop scalar mass correction shown in \fref{daisyresum} (a) scales as $\sim N \lambda T^2$ to leading order in temperature, which is the ``hard thermal loop''. The phase transition occurs around the temperature where this thermal mass cancels  the tree-level mass at the origin:
\begin{equation}
 \mu^2 \sim N \lambda T^2 \ \ \  \ \ \ 
\Rightarrow 
\ \ \ \  \ \ \ 
\alpha \  \equiv \ N \lambda \frac{T^2}{\mu^2} \ \sim \ 1
\end{equation}
 At $n$-loop order, the leading contribution in temperature to the thermal mass is given by daisy diagrams shown in \fref{daisyresum} (b): 
\begin{equation}
\delta m^2_{\text{n-loop daisy}}\sim \
   N^n
   \
    \frac{\lambda^n T^{2n-1}}{ \mu^{2n-3}} 
   \ \ \ , \ \ \ n \geq 2
\end{equation}
The ratio of the $n$ to the $n+1$ loop daisy contribution scales as 
\begin{equation}
\frac{\delta m^2_{n}}{\delta m^2_{n-1}} \sim \alpha
\end{equation} 
which is not parametrically small during the phase transition, causing the perturbative expansion to break down. This can also be understood as an IR divergent contribution (in the high $T$ limit) to the zero mode propagator.

To make the expansion more reliable, it is necessary to resum the the thermal mass by replacing the tree-level $m_\mathrm{tree}^2(\phi)$ in \eref{VefffiniteT} by $m^2(\phi) = m^2_\mathrm{tree}(\phi) + \Pi(\phi,T)$, where in the standard method, $\Pi$ is taken to be the leading contribution in temperature to the one-loop thermal mass. For scalars this can be obtained by differentiating $V_\mathrm{th}$ with respect to $\phi$: 
\begin{equation}
\Pi \sim \lambda T^2 + \ldots
\end{equation}
The ellipses represent subleading contributions in both the high-temperature expansion and coupling order, which are neglected.

This substitution automatically includes daisy contributions to all orders in the effective potential. The largest contributions which are not included are the two-loop ``lollipop'' diagrams shown in \fref{daisyresum} (c), scaling as $\lambda^2 T^2 N$,  and the three-loop superdaisy shown in \fref{daisyresum} (d), scaling as $\lambda^3 T^4 N^3/\mu^2$. Reliability of the perturbative expansion with the above thermal mass substitution requires
\begin{equation}
\label{e.beta}
\lambda \ll 1, \ \ \ \ \ \beta \equiv \frac{\lambda T N}{\mu} \ll 1 \ .
\end{equation}
These are obtained by requiring the ratio of the one-loop thermal mass to the sunset and the ratio of the two-loop daisy to the three-loop daisy to be small.

To illustrate how this resummation procedure is implemented in most BSM calculations, let us again turn to our $SM + N_S\times S$ benchmark model. The ``dressed'' effective potential is given by 
\begin{equation}
\label{e.VefffiniteTdressed}
V_\mathrm{eff}^{\mathrm{dressed}}(h,S,T) = V_0 + 
 \sum_i \left[ 
V^i_\mathrm{CW}(m_i^2(h,S) + \Pi_i) + V^i_\mathrm{th}(m_i^2(h,S) + \Pi_i,T)
\right]
\ ,
\end{equation}
where 
\begin{align}
\label{e.PiT2}
\Pi_h(0) &= \Pi_G(0) = T^2 \left(\frac{3}{16} g^2 + \frac{1}{16} {g'}^2 + \frac{1}{4} \lambda_t^2 + \frac{1}{2} \lambda + \frac{N_S}{12} \lambda_{HS}\right) 
\\
\nonumber\Pi_S(0) &= T^2 \left( \frac{1}{3}  \lambda_{HS} + \frac{N_S+2}{12} \lambda_{S}\right)\\
\nonumber \Pi^L_{GB}(0) &= \frac{11}{6} T^2  \ \mathrm{diag}(g^2, g^2, g^2, {g'}^2) 
\end{align} 
and $\Pi^L_{GB}$ is added only to the longitudinal gauge boson masses squared  in the gauge basis, which are then diagonalized. Gauge symmetry suppresses thermal contributions to the transverse mode~\cite{Espinosa:1992kf}. Note that fermions do not receive large thermal masses due to chiral symmetry protection. Furthermore, there are no zero modes $\omega_n=(2n+1)\pi T$, and as a result no IR divergences appear in the fermion propagator.

As we explain in \sref{formal}, substituting $m^2_\text{tree} + \Pi$ directly into the effective potential is called \emph{Full Dressing} (FD). Since the thermal mass $\Pi$ is explicitly evaluated only to leading order in the high-temperature expansion, we refer to this resummation procedure as \emph{Truncated Full Dressing} (TFD). TFD is the standard approach for BSM calculations of the EW phase transition. 

If $V_\mathrm{th}^i$ is expanded using the high-Temperature approximation of \eref{JBFhighT}, the field dependent terms in logs cancel cancel between $V_\mathrm{CW}^i$ and $V_\mathrm{th}^i$. The $y^2$ term gives an overall contribution proportional to $ T^2 \Pi_i$, which is field-independent when using only the leading-order contribution to $\Pi_i$ in temperature. 
This just leaves the $y^3$ term, which can be captured by adding $V_\mathrm{ring}^i$:
\begin{equation}
\label{e.VefffiniteTring}
V_\mathrm{eff}^{\mathrm{dressed}}(h,S,T) = V_0 + 
 \sum_i \left[ 
V^i_\mathrm{CW}(m_i^2(h,S)) + V^i_\mathrm{th}(m_i^2(h,S),T)
+
V_\mathrm{ring}^i(m_i^2(h,S),T)
\right]
\ ,
\end{equation}
where 
\begin{equation}
V_\mathrm{ring}^i(m_i^2(h,S),T) = - \frac{g_i T}{12 \pi} \left( \left[m_i^2(h,S) + \Pi_i\right]^{3/2} - \left[m_i^2(h,S)\right]^{3/2}
\right) \ .
\end{equation}
Adding $V_\mathrm{ring}^i$ amounts to resumming the IR-divergent contributions to the Matsubara zero mode propagator. It is tantamount to performing the $m_i^2 \to m_i^2 + \Pi_i$ replacement in the full effective potential, under the assumption that only the thermal mass of the zero mode matters, which is equivalent to making a high-Temperature approximation.

This is the version of the finite-temperature effective potential used in most BSM calculations. In some cases, \eref{VefffiniteTring} is used but with the full finite-temperature $V_\mathrm{th}^i$ instead of the high-T expansion. This is more accurate when $m_i$ is comparable to the temperature, but in that case the assumptions that justify using $V_\mathrm{ring}$ are explicitly violated, and \eref{VefffiniteTdressed} is the more consistent choice. In practice, there is not much numerical difference between these two recipes. As we discuss in \ssref{problems}, all of these TFD calculations have problems arising from using only the leading-order contribution of $\Pi_i$ in temperature.

\subsection{Types of Electroweak Phase Transitions}
\label{ss.EWPTtypes}

It is well-known that in the SM for $m_h \gtrsim 70 \gev$, the EWPT is not first-order~\cite{Bochkarev:1987wf, Kajantie:1995kf}. To make the PT first order, new physics effects have to be added to the SM to generate an energy barrier between two degenerate vacua at $T = T_c$. These BSM scenarios  can be broadly classified into a few classes (see also~\cite{Chung:2012vg, Contino:2016spe}) based on the origin of the barrier between the two degenerate vacua. These are phase transitions driven by thermal effects, tree-level renormalizable effects, loop effects at zero-temperature and  non-renormalizable operators. Note that our simple BSM benchmark model realizes the first three of these mechanisms. Rigorous study of the fourth mechanism will require the updated thermal resummation procedure we present in this paper.

\vspace{4mm}
\noindent \emph{PT driven by BSM Thermal Effects} \vspace{2mm}

It is possible that BSM bosonic degrees of freedom are present in the plasma. If they have the right mass and coupling to the SM Higgs, they can generate an energy barrier to make the PT strongly first order. Schematically, this can be understood as follows. If the boson(s) have tree-level mass $m_i^2 =  \mu_S^2 + \lambda_{HS} h^2$, the effective potential of \eref{VefffiniteTring} contains a term of the form
\begin{equation}
\label{e.cubicterm}
- \frac{T}{12 \pi}  \left[\mu_S^2 + \lambda_{HS} h^2 + \Pi_S\right]^{3/2} 
\end{equation}
If this term is dominated by the $h$-dependent piece at $T = T_c$, the resulting $\sim - \lambda_{HS} h^3 T$ negative cubic term can generate an energy barrier between two degenerate minima and catalyze a strong first order PT. 

In order for this cubic term to be manifest, it is required that $\mu_S^2 + \Pi_S \ll \lambda_{HS} v_c^2$. In the SM the $W$ and $Z$ bosons generate a cubic term, but their contribution is too small to make the SM EWPT first order. This can be enhanced in BSM scenarios by a partial cancellation between the new boson's thermal mass and a negative bare mass at $T = T_c$. 

This scenario was long regarded as one of the most promising avenues for EWBG, because light stops in supersymmetry could serve as these new bosonic degrees of freedom (DOF) \cite{Carena:1996wj, Laine:1998qk, Espinosa:1996qw, Delepine:1996vn, Carena:1997ki, Huber:2001xf, Carena:2002ss, Lee:2004we, Carena:2008rt, Carena:2008vj, Cirigliano:2009yd}.  Higgs coupling measurements have since excluded that possibility for the MSSM \cite{Curtin:2012aa, Cohen:2012zza} and general models with colored scalars \cite{Katz:2014bha, Katz:2015uja}.
Other scenarios, including the SM + $N_S \times S$ benchmark model we explore here, can easily realize this possibility~\cite{Katz:2014bha, Katz:2015uja}.

The mass of these light BSM bosonic DOF cannot significantly exceed $T_c \sim \mathcal{O}(100 \gev)$ to ensure their thermal contributions are unsuppressed. This makes such EWBG scenarios prime candidates for discovery at the LHC, and possibly future colliders. It is therefore of paramount importance to robustly correlate the predicted collider signatures with the regions of parameter space which allow for a strong phase transition. 

This mechanism relies on a partial cancellation between a zero-temperature mass and a thermal mass. However, in the standard calculation, the thermal mass is computed only to leading order in the high-T expansion. This is troubling, since (a) even within the high-T expansion, subleading terms in the expansion can change the thermal mass by $\mathcal{O}(40\%)$ or more \cite{Bellac:2011kqa}, and (b) the thermal mass should decrease for nonzero Higgs expectation values, since the bosons become heavier as $h \to v_c$ and partially decouple from the plasma. This can affect the electroweak phase transition, and the corresponding predictions for collider observables from a strong EWPT. Addressing this issue will be one of the major goals of our work.

\vspace{4mm}
\noindent \emph{PT driven by tree-level renormalizable effects} \vspace{2mm}

It is possible to add new scalars to the SM Higgs potential, see e.g. \cite{Profumo:2007wc, Jiang:2015cwa, Profumo:2014opa, Kotwal:2016tex}. In that case, the tree-level structure of the vacuum can be modified. For example, it is possible for the universe to first transition to a nonzero VEV of an additional singlet, only to transition to another vacuum with a nonzero Higgs VEV at a lower temperature. It is also possible for the Higgs to mix with new DOF  (i.e. both the Higgs and the new DOF acquire VEVs in our vacuum). In that case, the tree-level potential can have a barrier between the origin and the EWSB minimum, resulting in a strong one-step phase transition at finite temperature.

These tree-driven one- or two-step PTs can easily be very strongly first order, but can also cause runaway bubbles, which are incompatible with sufficient BAU generation \cite{Kozaczuk:2015owa}. On the other hand, the strong nature of these PTs might make them discoverable by future gravitational wave observations \cite{Grojean:2006bp}. It is therefore important to understand which regions of parameter space are associated with these types of phase transitions. 

An intriguing version of the two-step EWBG scenario is possible when a triplet scalar is added to the SM \cite{Patel:2012pi, Blinov:2015sna, Inoue:2015pza}. In that case, the baryon asymmetry can be created in the first transition to the triplet-VEV-phase, and preserved in the second transition to the doublet-VEV phase which the universe inhabits at zero temperature.

\vspace{4mm}
\noindent \emph{PT driven by loop effects at zero temperature} \vspace{2mm}

New degrees of freedom with sizable couplings to the Higgs can generate non-analytical contributions to $V_\mathrm{CW}$ at zero temperature which ``lift'' the local $h = v$ minimum to a higher potential relative to the origin, compared to the SM. With this shallower potential well, SM $Z$ and $W$ boson thermal contributions can be strong enough to generate a cubic potential term at finite temperature, resulting in a strong PT. This was recently discussed in the context of future collider signatures by~\cite{Curtin:2014jma}, and we will generalize their phenomenological results in this paper.

\vspace{4mm}
\noindent \emph{PT driven by non-renormalizable operators} \vspace{2mm}

The previous two phase transition classes are primarily associated with the zero-temperature effects of BSM degrees of freedom on the Higgs potential. If these states are sufficiently heavy, it might be reasonable to parametrize some of their effect in an EFT framework by adding a set or non-renormalizable operators to the SM Higgs potential. This was used to correlate Higgs self-coupling deviations with a strong EWPT~\cite{Grojean:2004xa, Bodeker:2004ws, Delaunay:2007wb}. 

While EFT analyses are useful for analyzing broad classes of new physics effects,  their construction and validity at finite temperature is not well-understood.\footnote{The authors of \cite{Damgaard:2015con} studied the agreement between a singlet extension of the SM and the corresponding EFT, but since TFD  was used decoupling effects could not be correctly modeled.} At zero-temperature experiments, like mono-energetic collisions with energy $E$, the effects of perturbatively coupled particle with mass $m$ can be well described by an EFT if $m/E > 1$. This is not the case in a plasma, where the heavy state can be directly excited even if $m/T$ is larger than unity, generating sizable thermal loop contributions. Furthermore, EFTs are problematic when studying phase transitions, since the spectrum which is integrated out changes between the two vacua. Finally, the agreement between a full theory including heavy states and an EFT description cannot presently be studied reliably. This is because in the TFD thermal mass resummation procedure, the effects of new particles in the full theory on light scalar thermal masses never decouples since $\Pi \sim T^2$ is independent of contributing particle masses. The non-decoupling of high mass DOFs in the full theory calculation is clearly unphysical, preventing us from understanding the EFT's radius of convergence in field space and temperature. This provides another strong motivation for treating thermal masses more carefully.

\subsection{Problems with the standard one-loop TFD calculation of the phase transition}
\label{ss.problems} 

There are a few ways in which the standard calculation with TFD thermal mass resummation, as outlined above, is incomplete and can be extended. 
\begin{enumerate}

\item \emph{Resumming Goldstones:} At zero temperature, SM Goldstone contributions must be resummed to eliminate the unphysical divergence in the derivatives of $V_\mathrm{CW}$ when their masses at tree-level are zero~\cite{Martin:2014bca, Elias-Miro:2014pca}. The numerical effects of the Goldstone contributions, once resummed, are small, so we can deal with this by not including Goldstones in the loop calculations of certain couplings. In the $\overline{\mathrm{MS}}$ scheme this is not a (numerical) problem as long as the tree-level Higgs VEV is somewhat shifted from the loop-level Higgs VEV.

\item \emph{Gauge dependence:} Since the potential is derived from the gauge-dependent 1PI effective action, $v_c$ is not a gauge-independent quantity. In the standard Landau-gauge-fixed calculation, we compute $v_c/T_c$ as a proxy for the sphaleron energy in the broken phase (which is gauge independent), and the requirement that $v_c/T_c > 0.6$ is understood to be an approximate minimal necessary condition for EWBG to be plausible. 

A fully  gauge-independent calculation of $T_c$ and the sphaleron energy would make the calculation more reliable. This problem was considered by the authors of~\cite{Patel:2011th} in the high-temperature approximation. The gauge dependent potential without any thermal mass resummation is
\begin{equation}
\label{e.VefffiniteTgauge}
V_\mathrm{eff}^{T>0}(h,S,T;\xi) = V_0 + 
 \sum_i \left[ 
V^i_\mathrm{CW}(m_i^2(h,S;\xi)) + V^i_\mathrm{th}(m_i^2(h,S;\xi),T)
\right]
\ ,
\end{equation}
where $\xi$ is the gauge parameter.
Consider the gauge dependence of the third term $V^i_\mathrm{th}$:
\begin{equation}\label{finiteTeffPot}
\begin{split}
V^i_\mathrm{th}=\frac{T^4}{2\pi^2}\Big[\sum_{\text{scalar},i}\!\!J_B\left(\frac{m_i^2(h,S;\xi)}{T^2}\right)+3\sum_{\text{gauge},a}\!\!J_B\left(\frac{m_a^2(h)}{T^2}\right)-\sum_{\text{gauge},a}\!\!J_B\left(\frac{\xi m_a^2(h)}{T^2}\right)\Big]\,,
\end{split}
\end{equation}
where $m_i^2(h;\xi)=m_i^2(h)+\xi m_a^2$, and we have dropped fermion contributions which do not contain any gauge dependence. In the high-temperature expansion and for small $\xi$, the $\xi$-dependent contribution of DOF $i$ charged under the gauge symmetry is
\begin{equation}
V^i_\mathrm{th}[\xi]=\frac{T}{2\pi^2}(m_i^3(h;\xi)-\xi m_a^3(h))+\frac{m_i^4(h;\xi)-\xi m_a^4(h)}{64\pi^2} \log\frac{\mu_R^2}{T^2} \ .
\end{equation}
Note that the $\mathcal{O}(T^2)$ term, and hence the dominant contribution to the thermal mass, is gauge-independent. This means that $T_c$ only has a small gauge dependence, confirmed by \cite{Patel:2011th} for small values of $\xi$. Furthermore, for singlet extensions the new contributions to the potential which drive the strong phase transition are by definition gauge-independent. Therefore we do not deal with the issue of gauge dependence here and proceed with the standard Landau gauge-fixed calculation. Certainly, further work is needed to construct a fully gauge-independent general calculation of the strength of the electroweak phase transition, and to understand how sensitive the results of a gauge-fixed calculation are to the choice of gauge parameter.

\item \emph{RG-improvement:} the convergence of the one-loop effective potential can be improved by using running couplings with 2-loop RGEs. This is independent of other improvements to the calculation and is most important when the theory contains sizable mass hierarchies. We will not discuss it further here.

\item \emph{Higher-loop corrections:} it is possible to evaluate higher-loop contributions to the zero- and finite-temperature effective potential, such as the 2-loop lollipop that is not included via thermal mass resummation. Alternatively, estimates of these contributions can be used to determine whether the one-loop expansion is reliable. In our BSM calculations we will carefully do the latter, using high-$T$ approximations for the relevant diagrams.

\item \emph{Consistent Thermal Mass Resummation:} In the standard Truncated Full Dressing calculation outlined above, the effective finite-temperature Higgs potential is computed by inserting the truncated thermal masses $\Pi_i \sim T^2$ into the one-loop potential as shown in \eref{VefffiniteTdressed}.  This is also called ``resumming hard thermal loops'', since it amounts to resumming only the contribution to the Matsubara zero mode propagator. This is indeed correct, if those contributions dominate the sum of diagrams, which is the case in the extreme limit of the high-temperature approximation. 

Early calculations that used this approximation~\cite{Gross:1980br, Parwani:1991gq, Arnold:1992rz, Carrington:1991hz} were interested mainly in the restoration of electroweak symmetry at high temperature. Determining $T_c$ with reasonable accuracy only requires considering the origin of the Higgs potential where the top and gauge boson masses are entirely dominated by thermal effects. In this case, the truncated high-$T$ expansion for $\Pi_i$ is justified, though there are significant deviations which arise from subleading terms in the high-temperature expansion even at the origin. 

However, when studying the strong first-order phase transition and computing $v_c$, we have to deal with finite excursions in field space which by definition are comparable to the temperature. For $h \sim T$, masses which depend on the Higgs VEV due to a Higgs coupling strong enough to influence the PT cease to be small at tree-level compared to thermal effects, and should start decoupling smoothly from the plasma. The resulting $h$-dependence of $\Pi_i(h,T)$ is therefore important. This is especially the case when the strong phase transition is driven by light bosons in the plasma, and therefore reliant on the partial cancellation between a zero-temperature mass and a thermal mass correction.
Obtaining correct collider predictions of a strong EWPT requires going beyond the TFD scheme.

As mentioned previously, the high-$T$ thermal mass resummation is also incompatible with any EFT framework of computing the electroweak phase transition, since in this approximation the contribution of heavy degrees of freedom to thermal masses does not decouple. This confounds efforts to find a consistent EFT description of theories at finite temperature. 
Since EFTs are such a powerful tool for understanding generic new physics effects at zero temperature, rigorously generalizing their use to finite temperature is highly motivated.

\end{enumerate}

We will concentrate on ameliorating the problems associated with TFD thermal mass resummation. Some of the necessary components exist in the literature. It is understood that a full finite-temperature determination of the thermal mass can give significantly different answers from the high-$T$ expansion for the thermal mass \cite{Bellac:2011kqa}. This was partially explored, to subleading order in the high-$T$-expansion, for $\phi^4$ theories ~\cite{Espinosa:1992gq, Espinosa:1992kf, Quiros:1992ez, Dine:1992wr, Boyd:1992xn, Boyd:1993tz}, but never in a full BSM calculation,  without high-temperature approximations. 

We will perform a consistent (to superdaisy order) finite-temperature thermal mass computation by numerically solving the associated gap equation and resumming its contributions  in such a way as to avoid miscounting important higher-loop contributions.  Since we are interested in the effect of adding new BSM scalars to the SM in order to generate a strong EWPT, we will be performing this procedure in the scalar sector only. 
We now explain this in the next section.

%%%%%%%%%%%%%%%%%%%%%%%%%%%%
%%%%%%%%%%%%%%%%%%%%%%%%%%%%
%%%%%%%%%%%%%%%%%%%%%%%%%%%%
\section{Formal aspects of finite-temperature mass resummation}
\label{s.formal}
%%%%%%%%%%%%%%%%%%%%%%%%%%%%
%%%%%%%%%%%%%%%%%%%%%%%%%%%%
%%%%%%%%%%%%%%%%%%%%%%%%%%%%

As outlined above, in a large class of BSM models a strong EWPT is generated due to new weak-scale bosonic states with large couplings to the Higgs. A near-cancellation between the new boson's zero-temperature mass and thermal mass can generate a cubic term in the Higgs potential, which generates the required energy barrier between two degenerate minima $h = 0, v_c$ at $T = T_c$. To more accurately study the phase transition (and correlated experimental predictions) in this class of theories, we would like to be able to compute the thermal masses $\Pi_i$ of scalars $i$ beyond the hard thermal loop approximation used in TFD. In other words, rather than resumming only the lowest-order thermal mass in the high-temperature expansion, $\Pi_i \sim T^2$, we aim to compute and resum the full field- and temperature-dependent thermal mass $\delta m_i(h, T)$, with individual contributions to $\delta m_i$ accurately vanishing as degrees of freedom decouple from the plasma. We also aim to formulate this computation in such a way that it can be easily adapted for other BSM calculations, and the study of Effective Field Theories at finite Temperature.

A straightforward generalization of the TFD calculation might be formulated as follows. The finite-temperature scalar thermal masses can be obtained by solving a one-loop gap equation of the form
\begin{equation}
\delta m_i^2 = \sum_j \partial^2_i (V_\mathrm{CW}^j + V_\mathrm{th}^j) \ ,
\label{e.schematicgap}
\end{equation}
where $\partial_i$ represents the derivative with respect to the scalar $i$. 
The hard thermal loop result $\delta m_i^2 \sim T^2$ is the solution at leading order in large $T/m$.
To obtain the finite-temperature thermal mass, we can simply keep additional orders of the high-T expansion, or indeed use the full finite-temperature thermal potential of Eqns.~(\ref{e.Vth}, \ref{e.JBF}) in the above gap equation. In the latter case, the equation must be solved numerically. Once a solution for $\delta m_i^2(h,T)$ is obtained, we can obtain the improved one-loop potential by substituting $m_i^2 \to m_i^2 + \delta m_i^2$ in $V_\mathrm{CW} + V_\mathrm{th}$ as in \eref{VefffiniteTdressed}.

Solving this gap equation, and substituting the resulting mass correction into the effective potential itself, is called \emph{Full Dressing} \cite{Espinosa:1992gq, Espinosa:1992kf, Quiros:1992ez}. This procedure is physically intuitive, but it is not consistent. Two-loop daisy diagrams, which can be important at $T = T_c$, are miscounted \cite{Dine:1992wr, Boyd:1992xn}. 

An alternative construction involves substituting $m_i^2 \to m_i^2 + \delta m_i^2$ in the first derivative of the effective potential. This tadpole resummation, called \emph{Partial Dressing}, was outlined for $\phi^4$ theories in \cite{Boyd:1993tz}. The authors claim that PD correctly counts daisy and superdaisy diagrams to higher order than FD.

There appears to be some confusion in the literature as to whether FD or PD is correct~\cite{Espinosa:1992gq, Espinosa:1992kf, Quiros:1992ez,Boyd:1993tz,Dine:1992wr, Boyd:1992xn}, but we have repeated the calculations of \cite{Boyd:1993tz}, and confirm their conclusions. Partial Dressing (a) consistently resums the most dominant contributions in the high-temperature limit, where resummation is important for the convergence of the perturbative finite-temperature potential, (b) works to higher order in the high-temperature expansion, including the important log term, to correctly model decoupling of modes from the plasma, and (c) is easily adaptable to general BSM calculations. We review the important features of PD below and then outline how to generalize this procedure to numerically solve for the thermal mass at finite temperature.

\subsection{Tadpole resummation in $\phi^4$ theories}

To explore the correct resummation procedure we first study a $\phi^4$ theory with $N$ real scalars obeying an ${O}(N)$ symmetry. This can then be generalized to the BSM theories of interest for EWBG. The tree-level potential is
\begin{equation}
V_0 = - \frac{1}{2} \mu^2 \phi_i \phi_i + \frac{1}{4} \lambda (\phi_i \phi_i)^2 \ .
\end{equation}
Without loss of generality, assume that all excursions in field space are along the $\phi_0$ direction.

Resummation of the thermal mass is required when high-temperature effects cause the fixed-order perturbation expansion to break down. We are therefore justified in using the high-$T$ expansion to study the details of the thermal mass resummation procedure and ensure diagrams are not miscounted. Conversely, when $T \sim m$, there is no mismatch of scales to produce large ratios that have to be resummed. In this limit, the thermal mass will be less important, but should decouple accurately, and the resummed calculation should approach the fixed-order calculation. We now review how the PD procedure outlined in \cite{Boyd:1993tz} achieves both of these objectives.

\subsubsection{Partial Dressing Results for $N = 1$}

\begin{figure}
\begin{center}
\includegraphics[width=1\textwidth]{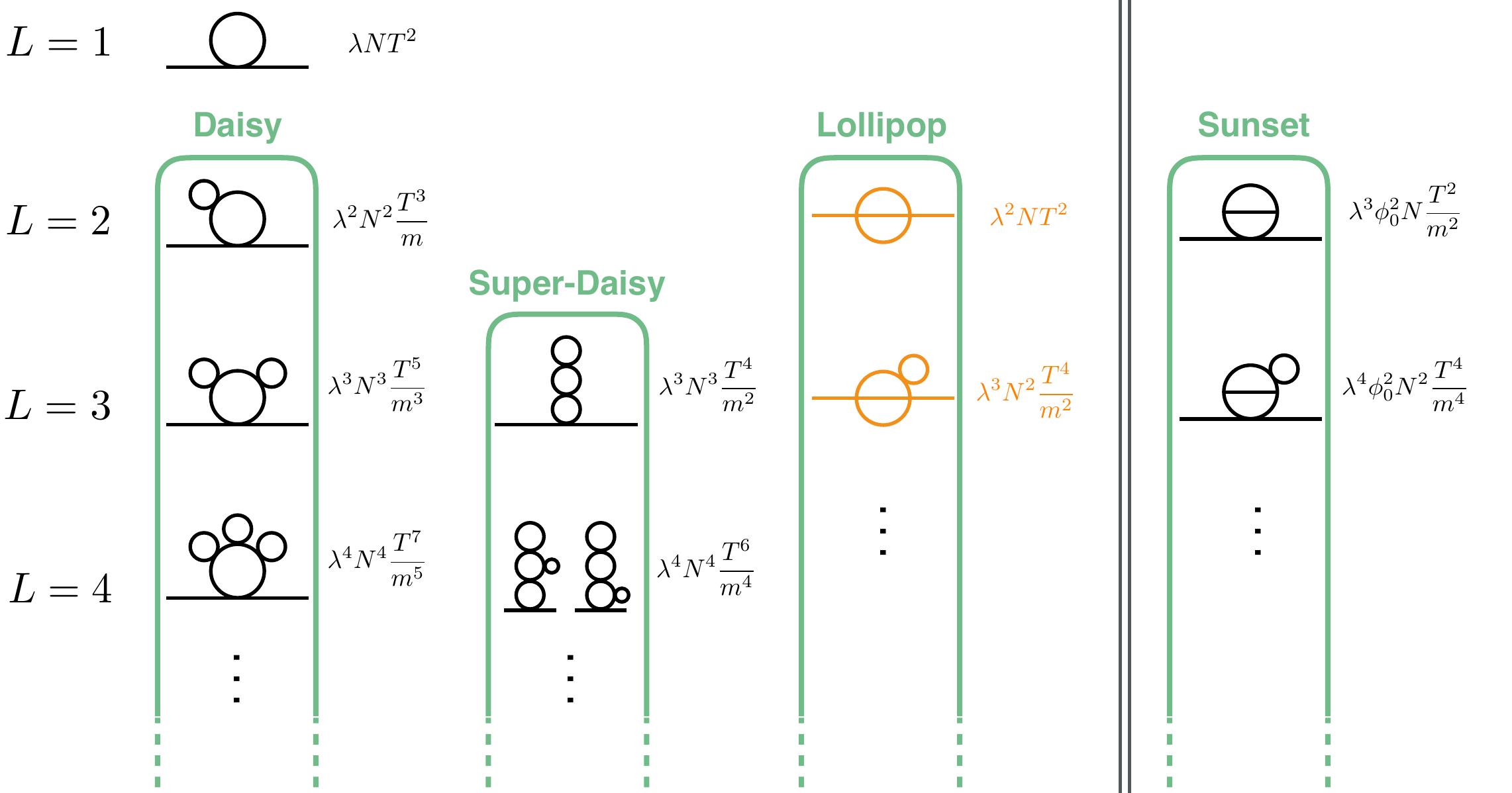}
\end{center}
\caption{
Complete set of 1- and 2- loop contributions to the scalar mass, as well as the most important higher loop contributions, in $\phi^4$ theory. The  scaling of each diagram in the high-temperature approximation is indicated, omitting symmetry- and loop-factors. 
Diagrams to the right of the vertical double-lines only contribute away from the origin when $\langle \phi \rangle = \phi_0 > 0$.  We do not show contributions which trivially descend from e.g. loop-corrected quartic couplings. Lollipop diagrams (in orange) are not automatically included in the resummed one-loop potential. 
}
\label{f.phi4diagrams}
\end{figure}

We start by summarizing the main result of \cite{Boyd:1993tz}, which studied the $N = 1$ $\phi^4$ theory in the high-temperature expansion.  
The first derivative of the one-loop effective potential $V_1 = V_\mathrm{CW} + V_\mathrm{th}$, see Eqns (\ref{e.CW0}), (\ref{e.Vth}) and (\ref{e.JBFhighT}), without any thermal mass resummation, is
\begin{equation} 
\frac{\partial V_1}{\partial \phi} \  = \ 
V_1^\prime 
\ = \ 
(6 \lambda \phi) \left[ \frac{T^2}{24} - \frac{T m}{8 \pi} - \frac{m^2 L}{32\pi^2}
\right]
\label{e.V1prime}
\end{equation}
where $L = \ln \frac{\mu_R^2}{T^2} - 3.9076$ and the log-term
arises from a cancellation between the zero- and finite-temperature potential. The tree-level scalar mass is $m^2 = -\mu^2 + \frac{\lambda}{2} \phi^2$. 
Differentiating this once again will yield the one-loop thermal mass shown in \fref{phi4diagrams}, as well as an electron-self-energy-type diagram for $\phi \neq 0$ that descends from loop corrections to the quartic coupling. During the phase transition, $\alpha \equiv N \lambda T^2/m^2 \approx 1$, requiring daisies to be resummed. This is evident in \fref{phi4diagrams} from the fact that subsequent terms in each family of diagrams (Daisy, Super-Daisy, Lollipop and Sunset) is related to the previous one by a factor of $\alpha$.

The second derivative of the one-loop potential defines a \emph{gap equation}, which symbolically can be represented as
\begin{center}
\includegraphics[width=0.8\textwidth]{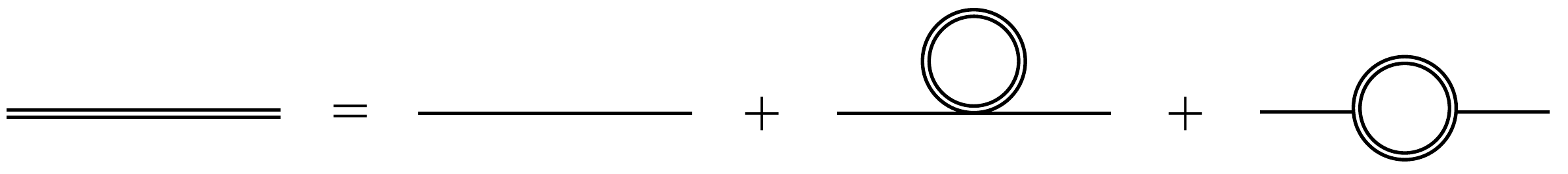}
\end{center}
where double-lines represent improved propagators with the resummed mass $M$, while single-lines are un-improved propagators with the tree-level mass $m$. Algebraically, this gap equation is obtained by substituting $M^2$ into the second derivative of the one-loop effective potential:
\begin{eqnarray}
\nonumber
M^2 &=& 
m^2 + \left.{V^{\prime \prime}_1}\right|_{m^2 \to M^2} 
\\ 
\label{e.phi4gapeqn}
\\
\nonumber
\Rightarrow  \ \ \ \ \ 
M^2  &=&
m^2 + \frac{\lambda T^2}{4} - \frac{3 \lambda T M}{4 \pi} 
- \frac{3 \lambda M^2 L}{16 \pi^2}
- \zeta \left[ 
\frac{9 \lambda^2 \phi^2 T}{4 \pi M}
 + \frac{9 \lambda^2 \phi^2}{8 \pi^2}
 \right]
\end{eqnarray}
where we have inserted a factor $\zeta = 1$ for reasons which will be made clear below.  The PD procedure involves resumming these mass corrections by substituting $m^2 \to M^2$ in the first derivative of the potential \eref{V1prime} rather than the potential itself. The potential is then obtained by integrating with respect to $\phi$:
\begin{equation}
\label{e.Vpd}
V_1^{pd} \equiv \int d\phi \left.V_1^\prime\right|_{m^2 \to M^2(\zeta = 1)} .
\end{equation}
By expanding the above in large $T$, one can show that $V_1^{pd}$ correctly includes all daisy and super-daisy contributions shown (in the form of mass contributions) in \fref{phi4diagrams}, to both leading, sub-leading and log-order in temperature. 

This partial dressing procedure does make one counting mistake, which is that all the sunset diagrams in \fref{phi4diagrams}, starting at 2-loop order and nonzero for $\phi \neq 0$, are included with an overall multiplicative pre-factor of $3/2$. This can be fixed by changing $\zeta$ in the gap equation (\ref{e.phi4gapeqn}) from 1 to $2/3$, resulting in the one-loop effective potential
\begin{equation}
\label{e.Vpd23}
V_1^{pd_{2/3}} \equiv \int d\phi \left.V_1^\prime\right|_{m^2 \to M^2(\zeta = 2/3)}
\end{equation}

Finally, non-daisy type diagrams, most importantly the two-loop lollipop in \fref{phi4diagrams} and its daisy-dressed descendants, are by definition not included in this one-loop resummed potential. However, in the high-temperature limit they can be easily included by adding the explicit expression for the lollipoop loop tadpole (one external $\phi$ line, hence the name),
\begin{equation}
\label{e.lollipop}
V_2^\prime \supset
V_\ell^\prime \ = \ (6 \lambda \phi) \frac{\lambda T^2}{32 \pi^2} \left[ \log \frac{m^2}{T \mu_R} + 1.65 \right]
\end{equation}
with the same $m^2 \to M^2(\zeta = 2/3)$ substitution:
\begin{equation}
\label{e.Vpdl23}
V_1^{pd_{2/3}+\ell} \equiv \int d\phi \left[V_1^\prime + V_\ell^\prime\right]_{m^2 \to M^2(\zeta = 2/3)}
\end{equation}
($\zeta = 1$ can also be used, in which case 2-loop sunsets are not corrected.)
This effective potential includes all daisy, superdaisy, sunset and lollipop contributions correctly, and is therefore correct to three-loop superdaisy order.

\subsubsection{Comparing resummation schemes}

We have verified that the above results generalize to $O(N)$ $\phi^4$ theories, with $N > 1$.
As mentioned above, at temperatures near the phase transition the parameter $\alpha \equiv N \lambda T^2/m^2$ is $\approx 1$, necessitating resummation. 
To compare different resummation approaches, let us first define which parameters are required to be small for the improved perturbative expansion to converge. Zero-temperature perturbation theory requires\begin{equation}
\label{e.Nlambdasmall}
N \lambda \ \ll \ 1
\end{equation}
(where the above equation, and other inequalities of its type, typically contain loop- and symmetry-factors which we usually suppress). Satisfying \eref{Nlambdasmall} means that during the phase transition, in regions of field space where $\alpha \approx 1$, the high-temperature expansion (whether in $T/m$ or $T/M$) is usually valid. 
In order for the series of high-temperature contributions to converge, the parameter $\beta$ must also be small:
\begin{equation}
\label{e.betasmall}
\beta \ \equiv \ N \lambda \frac{T}{m}  \ = \ \alpha \sqrt{N \lambda} \ \ll \ 1
\end{equation}
An easy way to see this is to examine \fref{phi4diagrams}. Let us call the mass contribution of the one-loop $\sim T^2$ diagram $\delta m_1$, and the total contributions of the daisy, super-daisy, lollipop and suset family of diagrams $\delta m^2_D,\delta m^2_{SD},\delta m^2_{LL},\delta m^2_{SS}$ respectively. After resummation in $\alpha$, we obtain\footnote{Note that in the individual diagrams of \fref{phi4diagrams}, the un-improved tree-level mass $m$ is used in the propagator. The entire e.g. lollipop series can be obtained by evaluating the leading diagram with the daisy-improved mass $M$.} 
\begin{eqnarray*}
\begin{array}{rclcrclcrcl}
\label{e.betaratio}
\delta m_1^2 &\sim& \lambda N T^2
&\ \ \ \ \  \ &
\delta m_D^2 &\sim& \lambda^2 N^2 \frac{T^3}{m}
&\ \ \ \ \  \ &
\delta m_{SD}^2 &\sim& \lambda^3 N^3 \frac{T^4}{m^2}
\\ \\
\delta m_{LL}^2 &\sim& \lambda^2 T^2 N
&\ \ \ \ \  \ &
\delta m_{SS}^2 &\sim& \lambda^3 \phi_0^2 N \frac{T^2}{m^2}
\end{array}
\end{eqnarray*}
Making use of $\alpha \approx 1$, these contributions arrange themselves in order of size:
\begin{eqnarray}
\frac{\delta m_D^2}{\delta m_1^2} \sim \frac{\delta m_{SD}^2}{\delta m_D^2}  &\sim& \alpha \frac{m}{T} = \beta
\\
\frac{\delta m_{LL}^2}{\delta m_D^2} &\sim& \frac{1}{N} \frac{m}{T} = \frac{1}{\alpha} \frac{1}{N} \beta
\\ 
\frac{\delta m_{SS}^2}{m^2_D} &\sim& \frac{\alpha}{N^2} \frac{\phi_0^2}{T^2} \frac{m}{T} = \frac{\phi_0^2}{T^2}\frac{1}{N^2} \beta
\end{eqnarray}
Clearly $\beta$ is the relevant expansion parameter which has to be small for the series to converge. 
Furthermore, in terms of $\beta$, both the lollipop and sunset diagrams are of the same order as the superdaisy family, but with additional suppression factors of $1/N$.\footnote{Due to the different symmetry factors of the lowest-order superdaisy and lollipop diagrams, the corresponding $N$-suppression is not numerically significant for $N \lesssim 10$.} (In our regime of interest, $\phi_0$ is usually not much larger than $T$, so the sunsets are subdominant or at most comparable to the lollipop and superdaisy.)

We can now carefully compare different resummation approaches. The partial dressing procedure, with the lollipop correction and the additional sunset contribution, is accurate to $\mathcal{O}(\beta^2)$. 
Since tree- and loop-contributions are of similar size near the phase transition we compare all sub-leading contributions to the unimproved one-loop thermal mass $\delta m_1^2$. Relative to $\delta m_1^2$, the size of neglected zero-temperature contributions and non-daisy contributions at three-loop order are
\begin{equation}
\label{e.pderrorterms}
(N \lambda)^2 \ \  \mathrm{and} \ \ \beta^3 \sim (N \lambda)^{3/2} \ ,
\end{equation}
respectively.

The alternative \emph{Full Dressing} procedure \cite{Espinosa:1992gq, Espinosa:1992kf, Quiros:1992ez} involves solving the same gap equation as for partial dressing, but substituting $m^2 \to M^2$ in the potential $V_1$ instead of its first derivative. This essentially dresses up both the propagator and the cubic coupling in the potential.\footnote{This inspires the name we use for the standard thermal mass resummation as reviewed in \ssref{standardresummation}. Since it involves  computing the mass correction to leading order in high temperature $\Pi \sim T^2$ and substituting $m^2 + \Pi$ into the effective potential, we call it \emph{Truncated Full Dressing}, even though at $\mathcal{O}(T^2)$ there is no actual difference between FD and PD.}

 The authors of \cite{Boyd:1993tz} demonstrate that FD miscounts daisies and super-daisies (starting at the 2-loop level), does not include sunset contributions, and includes lollipop contributions but with a wrong prefactor and without the log-dependence of \eref{lollipop}, which arises from neglecting internal loop momenta (as expected in a resummation procedure which does not explicitly calculate multi-loop diagrams). We have confirmed their results. Therefore, ignoring the incorrect accounting of the lollipop which vanishes at the origin, the error terms of a PD calculation are 
 \begin{equation}
 \label{e.fderrorterms}
(N \lambda)^2 \ \  \mathrm{and} \ \ \beta \sim (N \lambda)^{1/2}  \ ,
\end{equation}
 The standard BSM calculation is even worse, since TFD only uses the leading-order thermal mass, leading to possible error terms
\begin{equation}
\label{e.truncatedfderrorterms}
(N \lambda)^2 \ \  , \ \ \beta \sim (N \lambda)^{1/2} 
\ \ \mathrm{and} \ \ \frac{m}{T} \ .
\end{equation}

The advantages of partial dressing, compared to the truncated (or un-truncated) full dressing procedure, are clear, especially for phase transitions driven by BSM thermal effects, where the $m/T$ error in \eref{truncatedfderrorterms} can be significant.

\subsection{A general resummation procedure for BSM theories}
\label{ss.generalbsmcalc}

We now discuss how to adapt partial dressing for efficient calculation of the phase transition in general BSM theories. We will limit ourselves to phase transitions along the Higgs direction, briefly discussing other cases in the next subsection.

Since partial dressing avoids miscounting of the most important thermal contributions at all orders in the high-T expansion, including the log-term, it can be explicitly applied to the finite-temperature regime. 
In regions where $\alpha \approx 1$, the high-temperature expansion is valid, and resummation will be properly implemented. This smoothly interpolates to the regime where masses are comparable to temperature, eliminating the separation of scales and making the fixed-order calculation reliable again, with finite-temperature effects decoupling correctly as the mass is increased. Therefore, for a given set of mass corrections $\delta m_i^2$ for gauge bosons and scalars $i$, we define our effective potential along the $h$-direction by substituting the mass corrections into the first derivative of the loop potential:
\begin{equation}
\label{e.VefffiniteTdressedpp}
V_\mathrm{eff}^{\mathrm{pd}}(h,T) = V_0 + 
 \sum_i 
 \int dh
 \left[ 
\frac{\partial V^i_\mathrm{CW}}{\partial h}\Big(m_i^2(h) + \delta m^2_i(h, T)\Big) + \frac{\partial V^i_\mathrm{th}}{\partial h}\Big(m_i^2(h) + \delta m^2_i(h, T),T\Big)
\right]
\ ,
\end{equation}
Note that $V_\mathrm{th}$ is not (necessarily) expanded in high- or low-temperature. 

Next, how do we obtain the mass corrections $\delta m_i^2$? We will concentrate on cases, like our $SM + N_S\times S$ benchmark model, in which the dominant effect of new physics on the phase transition comes from an expanded scalar sector. Therefore, we will retain use of the $\mathcal{O}(T^2)$ gauge boson thermal masses of \eref{PiT2}, and set
\begin{equation}
\label{e.PiT2new}
\delta m_{GB}^2(h) = \Pi_{GB}(0)
\end{equation}
For the scalar mass corrections we numerically solve a set of coupled gap equations at each different value of $h$ and $T$:
\begin{equation}
\label{e.BSMgapeqn}
\delta m_{\phi_j}^2(h, T) =
\sum_i
 \left[ 
\frac{\partial^2 V^i_\mathrm{CW}}{\partial \phi_j^2}\Big(m_i^2(h) + \delta m^2_i(h, T)\Big) + \frac{\partial^2 V^i_\mathrm{th}}{\partial \phi_j^2}\Big(m_i^2(h) + \delta m^2_i(h, T),T\Big)
\right]
\end{equation}
where $\phi_j = h, G_0, S_0$. Since we only consider excursions along the $h$ direction, there are no mixed mass terms, and mass corrections for all singlets and Goldstones respectively are equal.\footnote{We can also evaluate the potential along the $S_0$ axis, in which case the mass corrections for $S_0$ and $S_k$ with $k \geq 1$ have to be treated separately. Note that \eref{BSMgapeqn} is also a function of the gauge boson masses and thermal masses, which are set by \eref{PiT2new}, as well as the fermion masses.}

Note that while we only numerically solve for the mass corrections of the scalars, these mass corrections will include contributions due to gauge bosons and fermions, which decouple correctly away from the high-temperature limit. We will address some subtleties related to finding consistent numerical solutions to these gap equations, and the effect of derivatives of $\delta m_{\phi_j}^2$, in the next section. 

In defining the effective potential \eref{VefffiniteTdressedpp}, we are essentially using only the one-loop potential as in \eref{Vpd}. This represents a great simplification, since calculation of the two-loop lollipop in full generality and at finite temperature~\cite{Parwani:1991gq} may be very onerous in a general BSM theory.  Furthermore, implementing the factor-of-$2/3$ ``fix'' to correctly count sunset contributions may be nontrivial at finite-temperature. Fortunately, we can show that omitting both of these contributions is justified for our cases of interest.

First, the lollipop is suppressed relative to the dominant one-loop resummed potential by factors of $N$ (where $N$ is related to $N_S$ but also the number of Goldstones in the SM) and $\beta < 1$. Even so, it represents our dominant neglected contribution. 
To explicitly check that it is small, it is sufficient to evaluate the dominant lollipop contributions to the $h$ and $S_0$ thermal masses in the high-temperature limit. Adapting the loop integral in the high-$T$ limit from \cite{Arnold:1992rz}, this gives
\begin{eqnarray}
\nonumber \delta m^2_{h, LL}(T)  &= & \frac{T^2}{16 \pi^2} 
\left[
{
6 \lambda^2 \left(\log \frac{m_h^2}{T \mu_R} + 1.65 \right)
+ 
N_S \lambda_{HS}^2 \left( \log \frac{(m_h + 2 m_S)^2}{9 T \mu_R} + 1.65 \right) 
}
\right]
\\
\label{e.deltamsqLL}
\\
\nonumber 
\delta m^2_{S, LL}(T)  &=&
\frac{T^2}{16 \pi^2}
\left[
{
(N_S + 2) \lambda_S^2 \left( \log \frac{m_S^2}{T \mu_R} + 1.65\right)
+
4 \lambda_{HS}^2 \left( \log \frac{(m_S + 2 m_h)^2}{9 T \mu_R} + 1.65\right)}
\right]
\end{eqnarray}
These diagrams are evaluated with improved propagators. In order for the calculation to be reliable, the ratios of lollipop to resummed one-loop mass corrections must satisfy
\begin{equation}
\label{e.lollipopratios} 
r_{LL}^{k} \ \  \equiv \ \  \frac{1}{\delta m_{k}^2(h=0,T_c)} \ 
\left.\delta m^2_{k, LL}(T_c)\right|_{m_h \to \sqrt{m_h^2 + \delta m_h^2}, m_S \to \sqrt{m_S^2 + \delta m_S^2}}  \ \  \lesssim \ \  0.1 \ ,
\end{equation}
for $k = h, S$ when the high-$T$ approximation is valid at the origin. (As explained in \ssref{finitetemperature}, in all operations involving the effective potential or its derivatives, we always only use the real part.)

Second, the sunset contribution is suppressed relative to the dominant one-loop-resummed potential by factors of $h/T$, $N^2$ and $\beta < 1$. More importantly, we expect the most important improvement of our partial dressing computation, compared to the standard truncated full dressing computation, to be the correct inclusion of finite-temperature effects, and the associated decoupling of heavy modes from the plasma away from the origin in field space. This decoupling will dominantly be due to the increased mass of the singlets as $h$ evolves away from the origin, rather than from the two-loop sunset contributions. Since the former is correctly captured by using finite-temperature gap equations and effective potential, the effect of sunsets should be small compared to the lollipop.

Third, we should also ensure that the equivalent of $\beta^3$ in our theory is sufficiently small at $T = T_c$. This is very straightforward using \eref{betaratio}, where $\delta m_1$ is the unimproved Higgs or Singlet one-loop thermal mass (using $m^2$ in propagators), and $\delta m_D^2$ is well approximated by the difference between the improved (using $M^2$) and un-improved thermal mass. Explicitly, we define two parameters:
\begin{equation}
\label{e.betahS}
\beta_k \equiv \frac{ | \delta m_k^2(0, T_c) -  \left.\delta m_k^2(0, T_c)\right|_{\delta m_i^2 \to 0}|}{\left.\delta m_k^2(0, T_c)\right|_{\delta m_i^2 \to 0}}
\end{equation}
for $k = h, S$. $\beta_h^3$ and $\beta_S^3$ then give the size of the error terms from the most important neglected three-loop diagrams and should be less than $\sim 0.1$. 

Finally,  two-loop zero-temperature corrections are small if $N \lambda \ll 4 \pi$, which simply restricts the weakly coupled parameter space we can explore. 
Our PD computation of the phase transition strength $v_c/T_c$ is then robust when all $r_{LL}^{h, S}$ and $\beta^3_{h,S}$ are small.

\subsection{Future Directions}

There are several conceptually straightforward ways to extend the procedure outlined in \ssref{generalbsmcalc}:
\begin{enumerate}

\item Even though it cannot be done by simple construction and manipulation of the one-loop effective potential, it would be straightforward to construct the gap equations for the gauge bosons~\cite{Espinosa:1992kf} and solve them together with the gap equations for the scalars.

\item With a general finite-temperature expression for the 2-loop lollipop shown in \fref{phi4diagrams}, we could evaluate its derivative, substitute $m^2 \to m^2 + \delta m^2$ and by adding it to \eref{VefffiniteTdressedpp} (but not \eref{BSMgapeqn}) correctly include all lollipop contributions, as in \eref{Vpdl23}

\item In regimes where the gap equation \eref{BSMgapeqn} can be approximated by a high-temperature expansion, one could implement the $2/3$ factor fix of \eref{phi4gapeqn} to correctly count sunset graphs at 2- and higher loop order. It is unclear how to implement this fix in the full finite-temperature gap equation, but as we discuss when we introduce \emph{Optimized Partial Dressing} (OPD) in 
\sssref{OPD}, the high-temperature expansion of the gap equation (but not the potential) is sufficient in most BSM calculations.

\item We have applied partial dressing without explicitly checking that fermion and gauge boson effects are correctly accounted for at 2- and higher-loop order. We leave this investigation to future work, but since the dominant BSM effects on the Higgs potential in our models of interest come from the scalar sector, we expect our procedure to be valid.

\end{enumerate}
Note that we explicitly check in our calculation whether the above-mentioned extensions (2) and (3) are numerically significant.

There is also a  more involved question, which is how to extend partial dressing for field excursions along several field directions at once, when those field directions cannot be related by a symmetry. 
 The partial dressing procedure unambiguously defines the potential along any one field direction, as long as all other fields are at the origin. In an example with two fields, let us denote as $V_\mathrm{eff}^{\mathrm{pd_i}}(\phi_1,\phi_2)$ the effective resummed one-loop potential obtained by integrating $\partial V_\mathrm{eff}/\partial \phi_i$. It is straightforward to show that $V_\mathrm{eff}^{\mathrm{pd_1}}(\phi_1,\phi_2) \neq V_\mathrm{eff}^{\mathrm{pd_2}}(\phi_1,\phi_2)$, with the difference being of super-daisy order $\mathcal{O}(\beta^2)$. This may not be numerically significant in a given case, but it would be of interest to extend the partial dressing procedure to consistently define 
\begin{equation}
V_\mathrm{eff}^{\mathrm{pd}}(\phi_1,\phi_2) \ .
\end{equation}
Since the above can be evaluated unambigiously diagrammatically, and since the partial dressing procedure was validated in \cite{Boyd:1993tz} by comparing this to the substitution of $M^2$ into the first derivative of the potential, there presumably exists a way of generalizing this substitution procedure to obtain a general potential as a function of multiple fields. This is of particular relevance to BSM models where the Higgs  mixes with BSM scalars \cite{Profumo:2007wc, Jiang:2015cwa, Profumo:2014opa}, which constitute an important class of models that can give a strong electroweak phase transition.

%%%%%%%%%%%%%%%%%%%%%%%%%%%%
%%%%%%%%%%%%%%%%%%%%%%%%%%%%
%%%%%%%%%%%%%%%%%%%%%%%%%%%%
\section{Computing the the Strength of the Phase Transition}
\label{s.calculation}
%%%%%%%%%%%%%%%%%%%%%%%%%%%%
%%%%%%%%%%%%%%%%%%%%%%%%%%%%
%%%%%%%%%%%%%%%%%%%%%%%%%%%%

We now apply the procedure outlined in \ssref{generalbsmcalc} the the SM + $N_S \times S$ BSM benchmark model with unbroken $O(N_S)$ symmetry in the $T = 0$ ground state.

\subsection{Zero-Temperature Calculation}
\label{ss.zeroT}

The tree-level scalar potential is given by Eqns.~(\ref{e.V0hS0}) and (\ref{e.Vcw}). All field excursions in the region of parameter space we study (no Higgs-Singlet mixing in the ground state) can be considered without loss of generality to be in either the $h$ or $S_0$ direction.

For each choice of $N_S$, there are three Lagrangian BSM parameters, $\mu_S, \lambda_{HS}$ and $\lambda_S$. We match these to three physical input parameters which are computed at one-loop level in the $\overline{\mathrm{MS}}$ scheme (in addition to matching the SM Higgs potential parameters $\lambda, \mu$ to $m_h$ and $v$):
\begin{enumerate}[label=(\alph*)]

\item The mass of the singlet in our vacuum 
\begin{equation}
m_S^2 = m_S^2(v) = \left.\frac{\partial^2 V_\mathrm{eff}^{T=0}}{\partial S_0^2}\right|_{h=v} = \mu_S^2 + \lambda_{HS} v^2 + \ldots
\end{equation}

\item The singlet-Higgs cubic coupling 
\begin{equation}
\lambda_{hSS}^\mathrm{loop} = \lambda_{hSS}^\mathrm{loop}(v) = 
\frac{1}{2} \left.\frac{\partial^3 V_\mathrm{eff}^{T=0}}{\partial S_0^2 \partial h}\right|_{h=v}
=
\lambda_{HS} v + \ldots
\end{equation}

\item The singlet quartic coupling
\begin{equation}
\lambda_{S}^\mathrm{loop} = \lambda_{S}^\mathrm{loop}(v) = 
\frac{1}{6} \left.\frac{\partial^4 V_\mathrm{eff}^{T=0}}{\partial S_0^4}\right|_{h=v}
=
\lambda_{S}  + \ldots
\end{equation}

\end{enumerate}
The renormalization scale $\mu_R$ is set to $m_S(v)$. In all calculations, we vary this scale choice up and down by a factor of 2 in order to estimate the uncertainty due to higher-order zero-temperature corrections. For a given set of input parameters, we compute whether  the singlet is stable at the origin when $h = 0$, and if not, the location of the local minimum $(h, S_0) = (0, w)$. When the singlet is unstable at the origin, we impose the vacuum stability condition 
\begin{equation}
 V_\mathrm{eff}^{T=0}(v,0) <  V_\mathrm{eff}^{T=0}(0,w)
 \end{equation}
to ensure the EWSB vacuum is the preferred one for our universe.

The collider phenomenology of the $N_S = 1$ case, with TFD resummation  and in the on-shell renormalization scheme, was studied previously in \cite{Curtin:2014jma}. 
The following three collider observables are of interest to probe this scenario:\footnote{We assume $3 \iab$ of luminosity at 100 TeV throughout.}
\begin{itemize}
\item A measurement of the triple-Higgs coupling with $5\%$ precision at $1 \sigma$, which is more pessimistic than recent estimates of the achievable precision~\cite{Contino:2016spe}.
\item A search in the VBF jets + MET channel for direct singlet pair production via $h^* \to S S$.
\item A measurement of the $Zh$ production cross section deviation from the SM with a $1 \sigma$ precision of 0.3\% at a TLEP-like lepton collider.
\end{itemize}
Together, it was found that these measurements provide coverage of the entire parameter space where a strong phase transition is possible. 

We analyze the same model as the authors of \cite{Curtin:2014jma}, generalized to $N_S \geq 1$, while comparing standard  truncated full dressing to our new partial dressing procedure. We therefore compute the same observables: 
\begin{enumerate}[label=(\alph*)]
\item The Higgs cubic coupling 
\begin{equation}
\label{e.lambdahhh}
\lambda_{hhh}^\mathrm{loop} = \lambda_{hhh}^\mathrm{loop}(v) = 
\frac{1}{6}
 \left.\frac{\partial^3 V_\mathrm{eff}^{T=0}}{\partial h^3}\right|_{h=v}
=
\frac{m_h^2}{2v} + \ldots
\end{equation}
To obtain the fractional deviation, we compare this to the value of  $\lambda_{hhh}^\mathrm{loop}$ computed without the $S$ contribution, rematched to the SM parameters  with the same choice of renormalization scale as the calculation with the singlet. 

\item Singlet pair production cross section at 100 TeV colliders: obtained in MadGraph~5~\cite{Alwall:2014hca} by using $\lambda_{hSS}^\mathrm{loop}$ as the singlet-Higgs tree-level coupling, scaled up by a factor of $N_S$.

\item The fractional $Zh$ cross section shift at lepton colliders can be computed using the results in ~\cite{Englert:2013tya, Craig:2013xia}:
\beq
\label{e.deltaZh}
\delta\sigma_{Zh}= \frac{N_S}{2}\frac{|\lambda_{HS}|^2 v^2}{4\pi^2 m_h^2}\left[1+F(\tau_\phi)\right]
\eeq
with the replacement $\lambda_{HS} v \to \lambda_{hSS}^\mathrm{loop}$. The loop function  $F(\tau)$, with $\tau_\phi=m_h^2/4m_S^2$, is given by
\beq
F(\tau)=\frac{1}{4\sqrt{\tau(\tau-1)}}\log\left(\frac{1-2\tau-2\sqrt{(\tau(\tau-1))}}{1-2\tau+2\sqrt{(\tau(\tau-1))}}\right).
\eeq

\end{enumerate}
Note that we estimate changes in cross-sections from the changes in potential couplings. This ignores finite-momentum contributions of new particles in loops to the cross section, but for $m_S \gtrsim m_h$, this is a good approximation.

Note that for these zero-temperature calculations, we do \emph{not} resum mass corrections in the partial dressing scheme. The procedures of \ssref{generalbsmcalc} can easily be applied to the zero-temperature potential as well, but we have extensively checked that they only have insignificant effects on the matched parameters and corresponding observables. This is expected, since mass resummation at finite temperature is required due to IR-divergent effects which are absent at zero temperature.\footnote{Note, however, that in the finite-temperature calculation, the zero-temperature potential gives important contributions beyond $T^2$ order.}

\subsection{Finite-Temperature Calculation}
\label{ss.finiteT}

We are interested in finding regions of parameter space where the phase transition is strongly one-step ($v_c/T_c > 0.6$) or two-step.

For each set of input parameters we match the potential, compute zero-temperature observables, and then compute the finite-temperature potential at different temperatures until we find $T =  T_c$ where the local minimum at the origin and a local minimum at $v = v_c$ are degenerate. 
The finite-temperature potential is given by \eref{VefffiniteTdressedpp}, where the mass corrections (or thermal masses) $\delta m_i^2$ are computed differently depending on the resummation scheme. 
If the singlet is unstable at the origin at $T = 0$,  we also compute the minimum temperature $T_S$ where thermal effects stabilize the singlet. If $T_S > T_c$, the transition is two-step and we do not analyze it further. 
In order for our calculation to be reliable, $r_{LL}^{h, S}$ and $\beta_{h,S}$ as defined in Eqns.~(\ref{e.lollipopratios}) and (\ref{e.betahS}), have to be small. 

We compare the three different schemes: Truncated Full Dressing (TFD), Partial Dressing (PD) and Optimized Partial Dressing (OPD). 
Some of the numerical calculations are performed in Mathematica, and some in a custom-built \texttt{c++} framework. For numerical evaluation, \texttt{c++} is operation-by-operation $\sim$ 1000 $\times$ faster than Mathematica, but the latter is much more versatile and often preferred for  exploring different BSM scenarios. Whenever evaluating the full thermal potential $V_\mathrm{th}$ or its derivatives without approximations, we use pre-computed lookup tables.

The TFD calculation, which is the standard method for BSM calculations used in \cite{Curtin:2014jma} and many other analyses, is simple enough that $v_c/T_c$ can be found for a given parameter point using about $\sim \mathcal{O}(1 \ \mathrm{minute})$ of CPU time in Mathematica. 
On the other hand, the complete implementation of PD is so numerically intensive that only the \texttt{c++} code can be realistically used, and evaluation times for a single parameter point range from 10 seconds to many minutes, indicating that PD is of order $10^4$ times more numerically intensive than TFD. 
Fortunately, a series of carefully chosen approximations allows partial dressing to be implemented in Mathematica, with only $\mathcal{O}(10\%$) higher  CPU cost than TFD, but identical results as PD. This is the OPD scheme, and we hope its ease of implementation and evaluation will be useful for future BSM analyses. 
We now briefly summarize and contrast the salient features of each resummation implementation.

% THREE CALCULATIONS:

\subsubsection{Truncated Full Dressing (TFD)}

This is just the standard thermal mass resummation, using $\delta m_i^2 = \Pi_i \propto T^2$, see \eref{PiT2}. This thermal mass is substituted into the first derivative of the potential, since there are no differences between full and partial dressing when substituting only the $T^2$ piece of the thermal mass. The thermal potential derivative $V_\mathrm{th}^\prime$ is evaluated numerically without any approximations. 

The dominant errors in the perturbative expansion arise from neglecting the two-loop lollipop, and miscounting the two-loop daisy.  We define the corresponding error term
\begin{eqnarray}
\label{DeltaTFD}
\Delta^\mathrm{TFD} &\equiv& \mathrm{max}(r_{LL}^h, r_{LL}^S, \beta_h, \beta_S)
\end{eqnarray}
to perform the perfunctory check that the calculation is reliable. 
However, is important to note that $\Delta^\mathrm{TFD}$ will greatly underestimate the error of the TFD calculation, since it does not include the $\mathcal{O}(m/T)$ errors in the truncated thermal mass.\footnote{In particular, the $m/T$ errors are expected to dominate  the $\beta$-size errors, since the differences between full and partial dressing, which are order $\beta$, only show up at subleading order in $T$.} It is precisely this error that will be explored by comparing the TFD calculation to (O)PD calculation.

\subsubsection{Partial Dressing (PD)}

This implements the scheme outlined in \ssref{generalbsmcalc} verbatim. For each temperature $T$ of interest, the algebraic gap equation \eref{BSMgapeqn} is constructed by substituting $m_i^2 \to m_i^2 + \delta m^2_i$ into the second derivative of the full finite-temperature thermal potential $V_\mathrm{th}^{\prime \prime}$ and $V_\mathrm{CW}^{\prime \prime}$. Solutions $\{ \delta m_i^2(h) \}$ are obtained numerically for each value of $h$. The resulting solutions $\delta m_i^2(h,T)$ are substituted into the first derivative of the full effective potential \eref{VefffiniteTdressedpp}, again without any approximations. The dominant errors arise from neglected lollipops and three-loop diagrams, and can be estimated with the error term
\begin{eqnarray}
\label{DeltaPD}
\Delta^\mathrm{PD} &\equiv& \mathrm{max}(r_{LL}^h, r_{LL}^S, \beta_h^3, \beta_S^3)
\end{eqnarray}

In practice, considerable complications arise when attempting to solve the gap equation \eref{BSMgapeqn}. We find the best possible solution (whether an exact solution exists or not) by minimizing $|\mathrm{Re}(\mathrm{LHS}) - \mathrm{Re}(\mathrm{RHS})|$ with respect to different choices of $\{\delta m_i^2\}$.
A unique solution to the gap equations can always be found at the origin of field space $(h,S) = (0,0)$.
However, sometimes no exact solution $\{\delta m_i^2\}$ can be found. This occurs for values of $h$ where some scalars become tachyonic, which is for example the case across the energy barrier. In that case, we use the closest approximate solution, or discard the solution and interpolate across these pathological values of $h$. Both methods  give very similar results, and in plots we use the former. The approximate solution usually still satisfies the gap equation at the $\sim$ 1 - 10 \% level, which on the face of it appears sufficient for a one-loop exact quantity.

Far more troubling is that for all other non-zero $h$-values, there exists not one but \emph{many}  numerical solutions to the algebraic gap equation. Apart from the computational intensity of PD, this is one of the confusing aspects which prompted us to develop OPD.\footnote{We have checked that our choice of solving a set of coupled gap equations for the scalar thermal masses, while using the analytical approximations for the gauge boson thermal masses \eref{PiT2}, is not responsible for either the absence or abundance of gap equation solutions at different values of $h$.}
Physically, one would expect the solution of the gap equation to correspond to the limit of an iteration, whereby propagators in diagrams contributing to the effective potential are recursively dressed with additional one-loop bubbles until their second derivative with respect to the field is consistent with the mass used in the propagators. 
Therefore, one way we attempt to ``find'' our way towards the physically relevant gap equation solution is by iteration, where the $n + 1$ step is given by
\begin{equation}
\label{e.gapeqniteration}
\delta m_{\phi_j}^2(h, T)_{n+1} =
\sum_i
 \left[ 
\frac{\partial^2 V^i_\mathrm{CW}}{\partial \phi_j^2}\Big(m_i^2(h) + \delta m^2_i(h, T)_n\Big) + \frac{\partial^2 V^i_\mathrm{th}}{\partial \phi_j^2}\Big(m_i^2(h) + \delta m^2_i(h, T)_n,T\Big)
\right]
\end{equation}
for fixed $h, T$. The iteration starts at $\delta m_{\phi_j}^2 = 0$ and continues until the result converges. 
However, this series does not always converge, instead oscillating between two or more values. This is always the case when there is no exact solution to the gap equation, but can also occur when there are one or more solutions. 
We numerically circumvent this issue by requiring the solution $\delta m_i^2(h,T)$ to be a set of smooth functions. Therefore, once solutions $\delta m_i^2(h,T)$ are obtained for a grid of $h$-values and fixed $T$, we use smoothing to eliminate numerical artifacts (or the jumps due to imperfect or multiple solutions), and interpolate across regions without solutions to obtain a smooth set of functions describing the thermal masses.

\subsubsection{Optimized Partial Dressing (OPD)}
\label{sss.OPD}

As we will discuss in \ssref{comparingschemes}, the results obtained via PD seem physically reasonable. However, it has two major disadvantages. First, its computational complexity and intensity would likely hamper adoption for other BSM calculations. Second, some of the numerical tricks used to obtain reasonable solutions to the gap equation are slightly unsatisfying. We are therefore motivated to develop the more streamlined OPD resummation scheme that is  numerically efficient and does not suffer from either the absence of gap equation solutions for some ranges of $h$ values, nor the preponderance of solutions for all other nonzero $h$ values. Encouragingly, the physical results of this OPD procedure are practically identical to PD. 

The first important observation, which is very well-known, is that the high-temperature approximation for the thermal potential, \eref{JBFhighT}, has the expected error terms of order $(m/T)^{n+1}$ if truncated at order $(m/T)^n$ (for $n \leq 2$), but is accurate for masses as large as $m \sim (1 - 3)\times T$  if the log terms are included. This applies both to $V_\mathrm{th}$ and its derivatives. 

Given that the low-temperature approximation \eref{JBFlowT} can easily be expanded to high enough order to be accurate for $m \sim T$, this implies that a piece-wise approximation for the thermal functions $J_{B,F}$ can be used in \eref{VefffiniteTdressedpp} to evaluate the effective potential, as well as its derivatives:
\begin{eqnarray}
\nonumber
J_B^\mathrm{piece-wise}(y^2) &=& \left\{
\begin{array}{ll}
J_B^{\mathrm{high}-T}(y^2)  & \mathrm{for} \ y^2 \leq 1.22
\\
\tilde J_B^{(3)}(y^2) & \mathrm{for} \ y^2 > 1.22
\end{array}
\right.
\\
\label{e.JBFpiecewise}
\\
\nonumber
J_F^\mathrm{piece-wise}(y^2) &=& \left\{
\begin{array}{ll}
J_F^{\mathrm{high}-T}(y^2)  & \mathrm{for} \ y^2 \leq 1.29
\\
\tilde J_F^{(2)}(y^2) & \mathrm{for} \ y^2 > 1.29
\end{array}
\right.
\end{eqnarray}	
This gives percent-level or better accuracy for $J_B$ and its first two derivatives and $J_F$ and its first derivative, for all positive $y^2$. For negative $y^2$ (corresponding to tachyonic masses) the accuracy is $\sim 10\%$ for $m^2 = -10 T^2$, but such negative $m^2$ are rarely encountered after thermal mass corrections are added. Evaluation of this piece-wise approximate form of $J_{B,F}$ is very fast in Mathematica.

This definition of the thermal effective potential $V_\mathrm{th}$ also allows the algebraic gap equation \eref{BSMgapeqn} to be defined entirely analytically (as opposed to numerically), even if the ultimate solutions have to be found numerically. This represents a huge simplification and allows Mathematica to find solutions much easier than for the full $V_\mathrm{th}$ defined via lookup-tables. In its full piece-wise defined form, the thermal functions of \eref{JBFpiecewise} still allow for the study of e.g. decoupling effects as particles become heavy, which will be important in the future study of EFTs at finite temperature. 

However, for the study of strong phase transitions induced by BSM thermal effects, we can make another important simplification. Thermal mass resummation is only needed to obtain accurate results when $m \lesssim T$. Therefore, we are justified in constructing the gap equation \eref{BSMgapeqn} using \emph{only the high-temperature approximation for the thermal potential} (and the usual $V_\mathrm{CW}$). The resulting solutions for the mass corrections $\{\delta m_i^2(h,T)\}$ are practically identical to the solutions obtained with the full finite-temperature potential, except for some modest deviations in regions where $m \gtrsim T$. Even so, the resulting effective potential obtained by integrating \eref{VefffiniteTdressedpp} is practically identical in those regions as well, since thermal effects of the corresponding degrees of freedom are no longer important. Using only $J_{B,F}^{\mathrm{high}-T}$ in the gap equation makes finding solutions so fast that the associated computational cost becomes a subdominant part of the total CPU time required for finding $v_c/T_c$, making this OPD method only  $\mathcal{O}(10\%)$ slower than the standard TFD method (even when the TFD method uses the same piece-wise defined thermal functions in the effective potential).

\vspace{3mm}
\noindent \emph{Alternative formulation of the gap equations}
\vspace{2mm}

\noindent The absence of solutions to the \emph{algebraic} gap equation \eref{BSMgapeqn} encountered in PD indicate an overconstrained system, meaning there might be missing variables we should also solve for. Furthermore, the numerical tricks utilized in the PD implementation to obtain $\delta m_i^2 (h,T)$ solutions for a given $T$ were justified by appealing to the required continuity of the solution (interpolation, smoothing) and the physical interpretation of the gap equation (selecting the correct solution by guiding the numerical root-finding procedure with iteration). 

All of these considerations point towards a slightly modified form of the gap equation which appears more consistent with the partial dressing procedure. Recall that the gap equation for both the full and partial dressing procedures, \eref{phi4gapeqn}, was originally defined by substituting $m^2 \to M^2$ in $V^{\prime \prime}$ of the $\phi^4$ theory:
\begin{equation}
\label{e.phi4gapeqnduplicate}
M^2 = m^2 + 
\left.{V^{\prime \prime}_1}\right|_{m^2 \to M^2} 
\end{equation}
This yields the gap equation used in the PD procedure \eref{BSMgapeqn}, which is an algebraic equation that is solved for $\{\delta m_i^2 \}$, a priori separately for each $(h,T)$. 
Alternatively, one could define the gap equation by taking the effective potential of partial dressing $V_1^{\prime pd} = \left.{V^{\prime}_1}\right|_{m^2 \to M^2} $ and differentiating it once more with respect to the field:
\begin{equation}
\label{e.phi4gapeqn2}
M^2 = m^2 + 
\left[\left.{V^{\prime}_1}\right|_{m^2 \to M^2} \right]^\prime
\end{equation}
In the context of our BSM benchmark model, the corresponding gap equations are
\begin{equation}
\label{e.BSMgapeqn2}
\delta m_{\phi_j}^2(h, T) =
\sum_i
\frac{\partial}{\partial \phi_j}
 \left[ 
\frac{\partial V^i_\mathrm{CW}}{\partial \phi_j}\Big(m_i^2(h) + \delta m^2_i(h, T)\Big) + \frac{\partial V^i_\mathrm{th}}{\partial \phi_j}\Big(m_i^2(h) + \delta m^2_i(h, T),T\Big)
\right]
\end{equation}
At each point in $(h, T)$ space, these gap equations are algebraic relations of $\{\delta m_i^2 \}$ as well as the derivatives at that point $\{ \frac{\partial \delta m_i^2}{\partial \phi_j}\}$. The gap equations are now \emph{partial differential equations}: The additional variables (the derivatives of $\delta m_i^2$) guarantee that solutions exist at every point\footnote{Symmetry under $\phi_i \to - \phi_i$ implies that $\frac{\partial \delta m_i^2}{\partial \phi_j} = 0$ when $\phi_j = 0$. Since we only consider excursions along the $h$-direction, only the derivatives with respect to $h$ are ever nonzero. At the origin both kinds of gap equations are the same, but at that point the algebraic gap equation always has a unique solution anyway.}, while the continuity condition of the PDEs
\begin{equation}
\label{e.continuitycondition}
\delta m_i^2(h+ \Delta h, T) = \delta m_i^2(h, T) + \Delta h \frac{\delta m_i^2}{\partial h}(h,T) + \mathcal{O}(\Delta h^2)
\end{equation}
restricts the number of numerical solutions, selecting the unique physical solution. This eliminates both numerical problems of the original PD procedure.

In practice, constructing \eref{BSMgapeqn2} is very challenging in the full finite-temperature formulation of $V_\mathrm{th}$, since it involves second derivatives of the thermal functions which are very computationally costly to evaluate. However, in the high-temperature approximation\footnote{One could also use the piece-wise definition of the thermal functions \eref{JBFpiecewise}, but obtaining a solution takes $\mathcal{O}(10)$ times longer. This can still be useful for studying decoupling effects and matching to EFTs.} \eref{BSMgapeqn2} becomes a very simple analytical set of equations that Mathematica can easily solve numerically for $\{\delta m_i^2\}$ after eliminating the derivatives $\{\frac{\partial \delta m_i^2}{\partial h}\}$ via the continuity condition \eref{continuitycondition}, making the solution at $(h + \Delta h, T)$ dependent on the solution obtained at $(h, T)$.  The resulting $(h,T)$-dependent mass corrections $\delta m_i^2(h,T)$ are well-behaved, defined everywhere, and unique almost everywhere.\footnote{In some regions there are multiple near-degenerate solutions, but the ambiguity is not physically significant.}

In summary,  the OPD method uses the piece-wise defined thermal functions \eref{JBFpiecewise} in the effective potential \eref{VefffiniteTdressedpp}, and the  high-temperature approximation in the gap equations \eref{BSMgapeqn2}. The gap equations are PDEs instead of simple algebraic relations, which leads to solutions for the mass corrections which are continuous and well-defined for all $(h, T)$ in the regions of interest.

Using gap equation \eref{phi4gapeqn2} instead of \eref{phi4gapeqnduplicate} amounts to treating resummation identically in the effective potential and the gap equation derived from that potential. While this seems reasonable, the important question is whether this particular \emph{algebraic} procedure of constructing a gap equation by manipulating the one-loop potential expressions, and inserting the resulting mass solution back into the potential, is equivalent (up to some order) to the \emph{diagrammatic} procedure of computing various higher-order contributions to the effective potential. 
As we reviewed in \sref{formal}, the authors of \cite{Boyd:1993tz} showed that using the original gap equation \eref{phi4gapeqnduplicate}, this equivalence was accurate up to differences of $\mathcal{O}(\beta^3)$, the neglected two-loop lollipop, and the miscounted two-loop sunset, see \fref{phi4diagrams}. 
We have checked that using the new gap equation \eref{phi4gapeqn2} is equivalent to the same order, the only difference being the precise nature of how the two-loop sunset is miscounted. 
This means that the $2/3$ correction factor in \eref{Vpdl23} is modified at subleading (i.e. 3-loop sunset) order.

In \ssref{generalbsmcalc} we argued that the sunset error is subdominant to the $\mathcal{O}(\beta^3)$ and lollipop errors, which we check explicitly are small. 
We therefore expect the numerical difference between $\delta m_i^2(h,T)$ solutions obtained in the OPD and PD scheme to be even smaller, and as we show below, this is indeed the case. That being said, since the gap equation in the OPD scheme is explicitly written in the high-temperature approximation, inserting the $2/3$ correction factor would be very straightforward.

Finally, one might also worry that the piece-wise defined thermal potential, or the high-temperature approximate gap equation, would be of insufficient accuracy in regions where large cancellations are important, or where masses are very tachyonic and the piecewise approximation is not very accurate. However, thermal resummation prevents masses squared inserted in the thermal potential from being too negative, and in regions where the transition between the high- and low-temperature approximation occurs, the exact solution to the gap equation becomes less numerically relevant since the calculation reduces to a fixed-order one.

\subsection{Comparing Resummation Schemes}
\label{ss.comparingschemes}

\begin{figure}
\begin{center}
\hspace*{-8mm}
\begin{tabular}{c}
\includegraphics[width=0.5\textwidth]{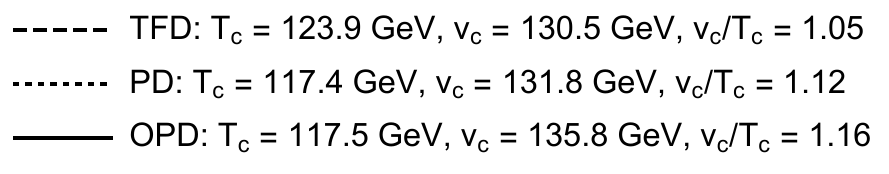}
\\ \\
\includegraphics[height=5cm]{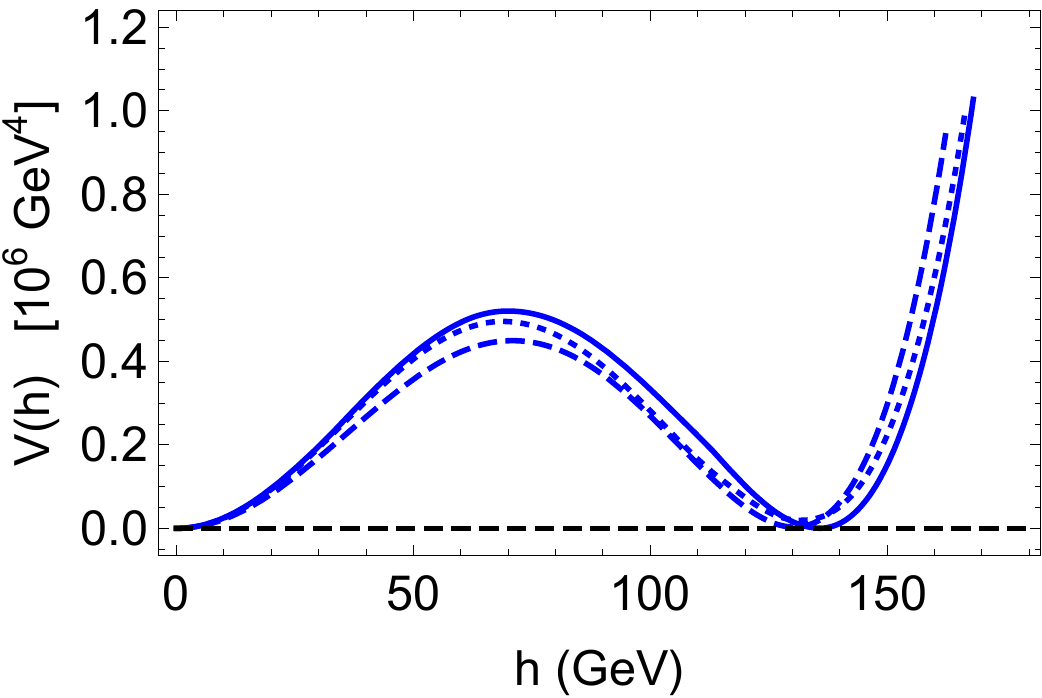}
\phantom{bla}
\includegraphics[height=5cm]{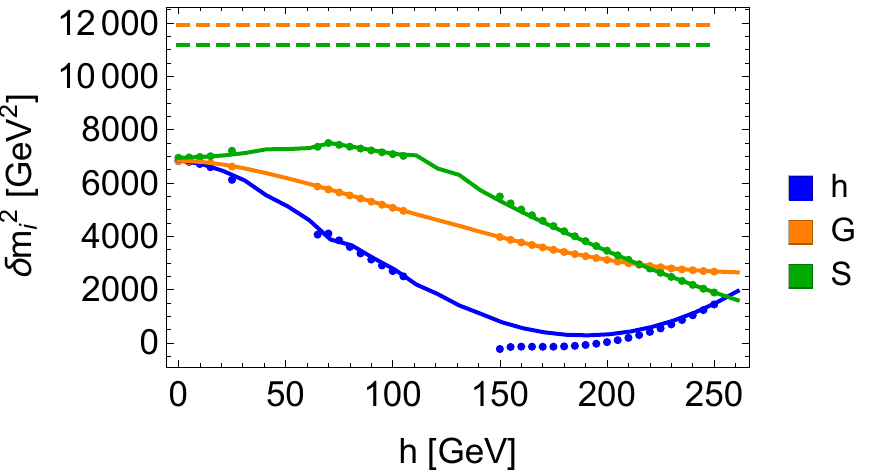}
\end{tabular}
\end{center}
\caption{
Effective Higgs Potential (left) and mass corrections $\delta m_i^2$ (right) for the physical Higgs ($h$), Goldstones ($G$), and singlets ($S$) at $T = T_c$ as a function of $h$. Evaluated in TFD, PD and OPD resummation schemes 
for $N_S = 3$ and
$(m_S, \lambda_{hSS}^\mathrm{loop}/v, \lambda_{S}^\mathrm{loop}) = (300 \gev, 1.52, 0.5)$.
In the right plot, $\delta m_h^2 = \delta m_G^2$ in the TFD scheme. The dots correspond to $\delta m_i^2$ in the PD scheme, with gaps indicating regions of the $h$-axis where no exact solution to the gap equation can be found, and the $\delta m_i^2(h, T)$ functions used to evaluate the potential are obtained by linearly interpolating between the obtained $\delta m_i^2$ solutions as a function of $h$. This gives nearly the same $V(h, T_c)$ as OPD. Note that the approximate equality of the three (O)PD mass corrections at the origin is a numerical coincidence for this parameter point. Furthermore, the differences in $v_c$, $T_c$ between TFD and (O)PD are modest here, but for other choices they can be much more pronounced. This is very important when $T_c \sim T_S$ and the predicted nature of the transition can change from one-step to two-step, as we discuss in \sref{physical}.
}
\label{f.comparisonplot}
\end{figure}

Here we illustrate the differences between the three resummation schemes for an example point in the parameter space of the SM + $N_S \times S$ model. 
The differences, in particular of the evaluated thermal mass corrections at $T = T_c$, will inform our discussion of the different physical predictions generated by the new (O)PD vs the standard TFD scheme in the next section. It will also demonstrate that PD and OPD are nearly equivalent.  

We focus on regions of parameter space where both TFD and (O)PD produce a sizable first-order one-step phase transition. This allows us to show the resulting Higgs potentials at $T = T_c$ side-by-side, but by necessity restricts our attention to regions of parameter space where all calculation schemes give similar physical predictions. Even so, the differences are very clear in detail and allow us to understand the regions of parameter space where the physical differences are more significant. 

\fref{comparisonplot} shows the effective Higgs potential and mass corrections for the parameter point $(N_S$, $m_S$, $\lambda_{hSS}^\mathrm{loop}/v$, $\lambda_{S}^\mathrm{loop}) = (3, 300 \gev, 1.52, 0.5)$. Immediately we see that PD and OPD give nearly identical functional forms of the mass corrections $\delta m_{h,G,S}^2(h,T_c)$ once OPD solutions are interpolated. As a result, the effective potential and obtained values of $v_c$ and $T_c$ are also nearly identical. We have checked that this holds true across the parameter space. Therefore, the numerically extremely efficient OPD method can be used in place of the numerically costly PD calculation.\footnote{For one-step transitions with very large couplings, where the zero-temperature singlet mass is small near the origin, but significantly larger than the Higgs mass at $h = v$, there are minor $\mathcal{O}(10\%$) differences between OPD and PD due to the assumed high-$T$ approximation in the gap equation. However, these differences do not significantly affect the physically important boundaries between different phases of the theory, where the PT is strongly first order, with one-step, or two-step transitions.}

The mass corrections obtained with (O)PD behave in a physically reasonable manner, being maximal near the origin and generally decreasing as $h$ increases and the various degrees of freedom acquire more mass, reducing their participation in the thermal plasma. The physical Higgs and Goldstone mass corrections behave differently away from the origin. Compared to the constant TFD prediction, the (O)PD  thermal masses are smaller by $\mathcal{O}(40\%$) or more. This is typical across the whole parameter space, and explains the most important physical difference between the two schemes. The reduced thermal masses in TFD result in higher temperatures $T_S$ where the singlet is stabilized at the origin (if it is unstable at zero temperature). As we show in \sref{physical}, this results in larger regions of parameter space where a two-step transition occurs.

%%%%%%%%%%%%%%%%%%%%%%%%%%%%
%%%%%%%%%%%%%%%%%%%%%%%%%%%%
%%%%%%%%%%%%%%%%%%%%%%%%%%%%
\section{Physical Consequences}
\label{s.physical}
%%%%%%%%%%%%%%%%%%%%%%%%%%%%
%%%%%%%%%%%%%%%%%%%%%%%%%%%%
%%%%%%%%%%%%%%%%%%%%%%%%%%%%

From a formal point of view, development of the (O)PD thermal resummation scheme is most important in the careful study of thermal decoupling effects, especially when spectra change with field excursions. This is necessary for rigorously understanding Effective Field Theories at finite temperature. We are currently pursuing this line of investigation, and will present the results in a future publication.   Additionally, compared to TFD calculations, the new (O)PD formalism makes quantitatively different predictions about the regions of parameter space where a one- or two-step phase transition of sufficient strength for EWBG can occur.   This in turn affects the predictions of the EWPT (or other phase transitions in the early universe) for cosmological observations and collider experiments which need to be known reliably for planning such experiments.  In more realistic extensions of the SM scalar sector, where the individual masses and couplings are not free parameters as in the SM + $N_S \times S$ benchmark model, this could also affect whether a strong phase transition is possible at all.

\begin{figure}
\begin{center}
\hspace*{-7mm}
\begin{tabular}{cc||cc||c}
\includegraphics[width=0.3\textwidth]{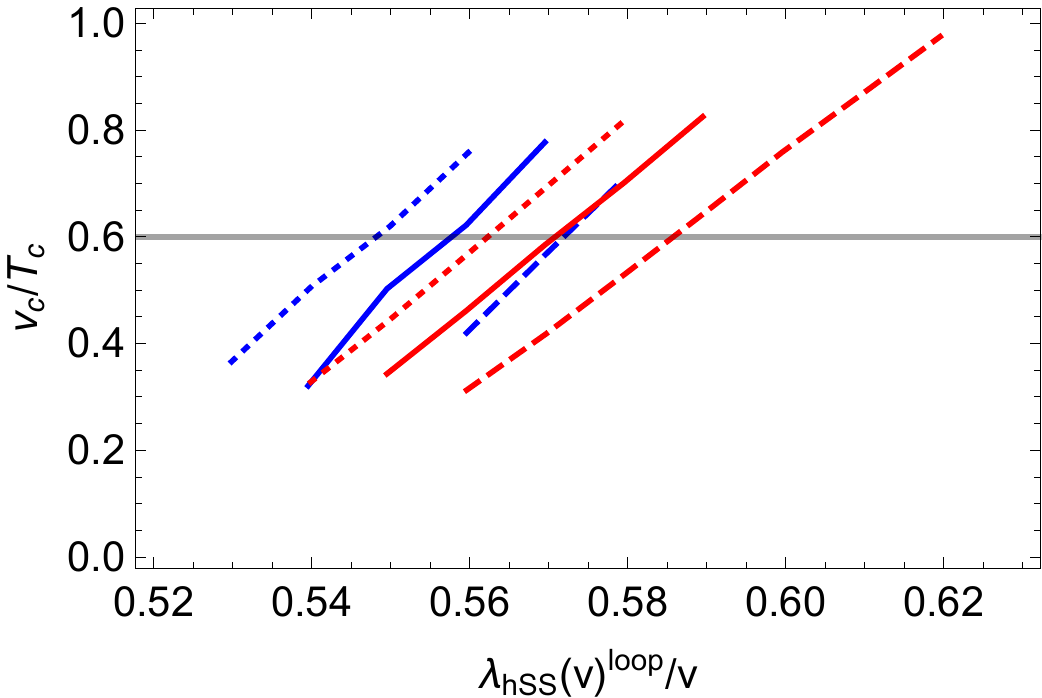}
&&
\includegraphics[width=0.3\textwidth]{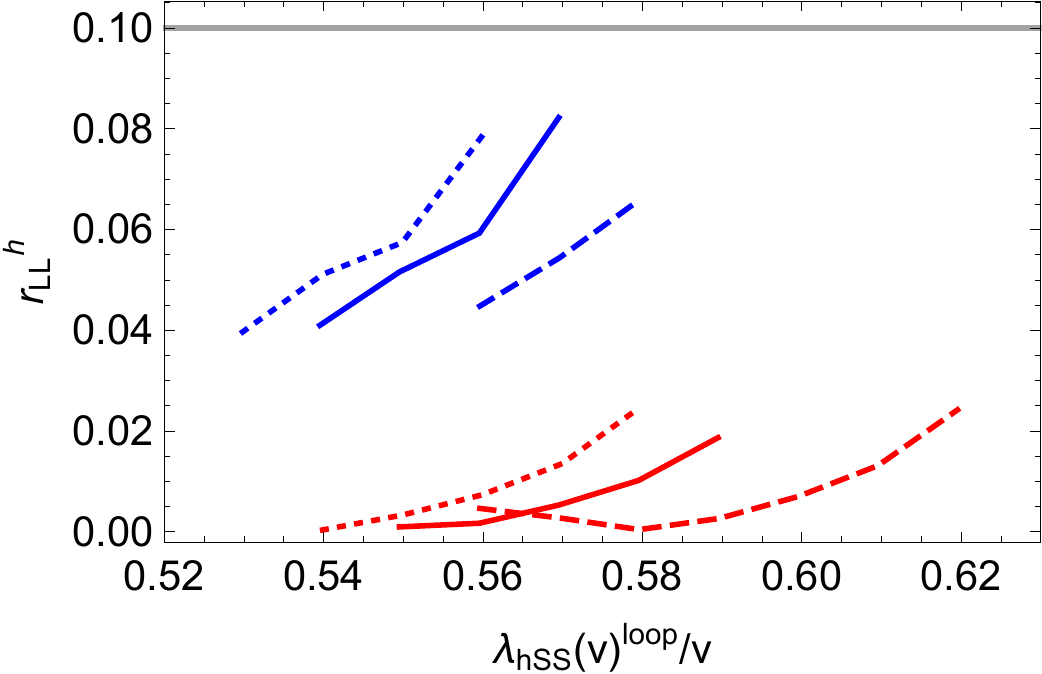}
& & 
\includegraphics[width=0.3\textwidth]{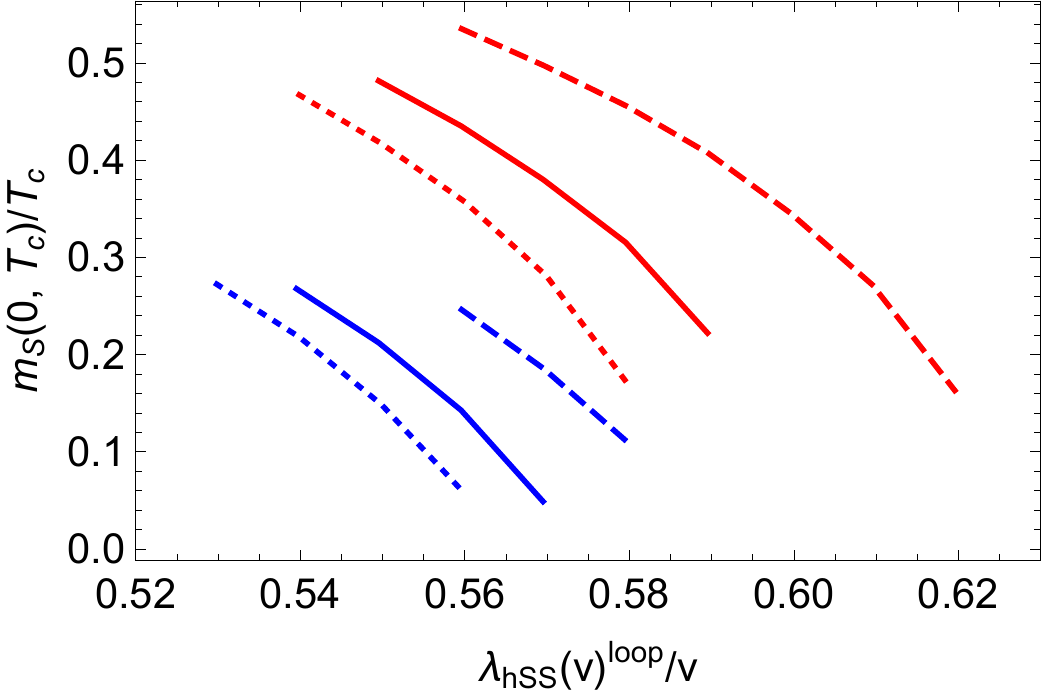}
\\
\footnotesize (a) $v_c/T_c$ && \footnotesize (e) $r_{LL}^h$ && \footnotesize (i) $m_S$ at origin\\  &&&&\\
\includegraphics[width=0.3\textwidth]{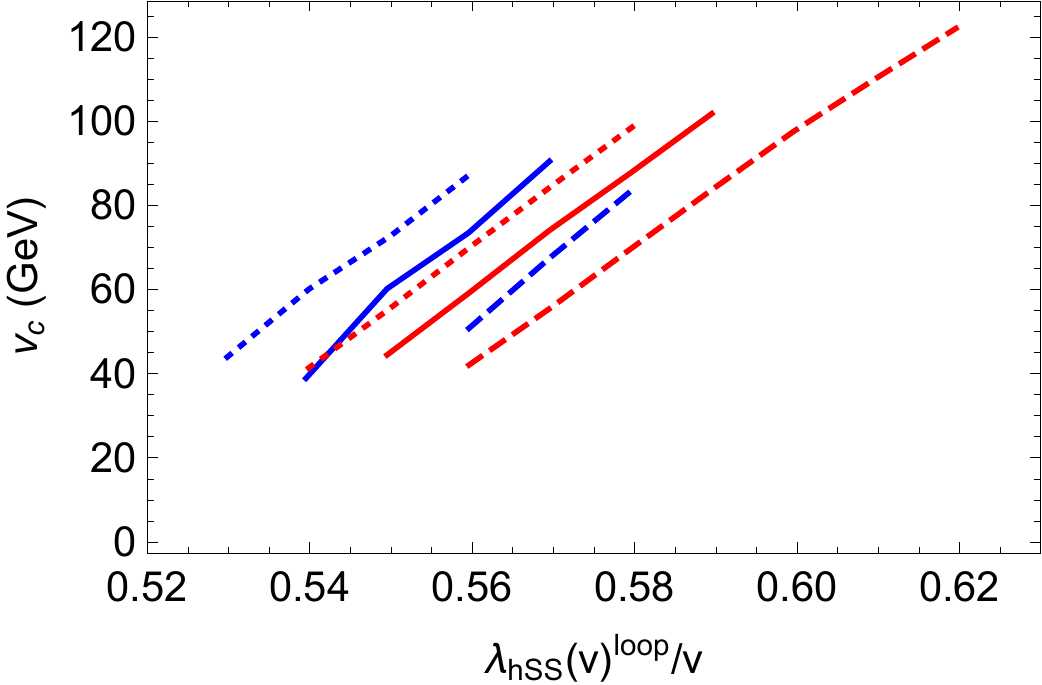}
&&
\includegraphics[width=0.3\textwidth]{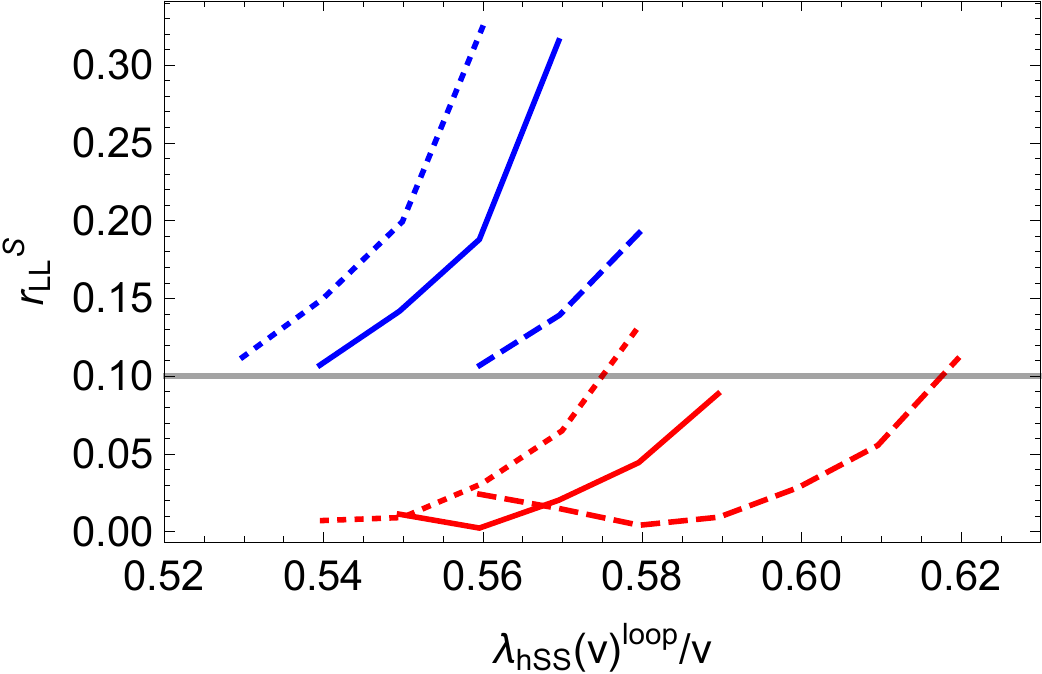}
& & 
\includegraphics[width=0.3\textwidth]{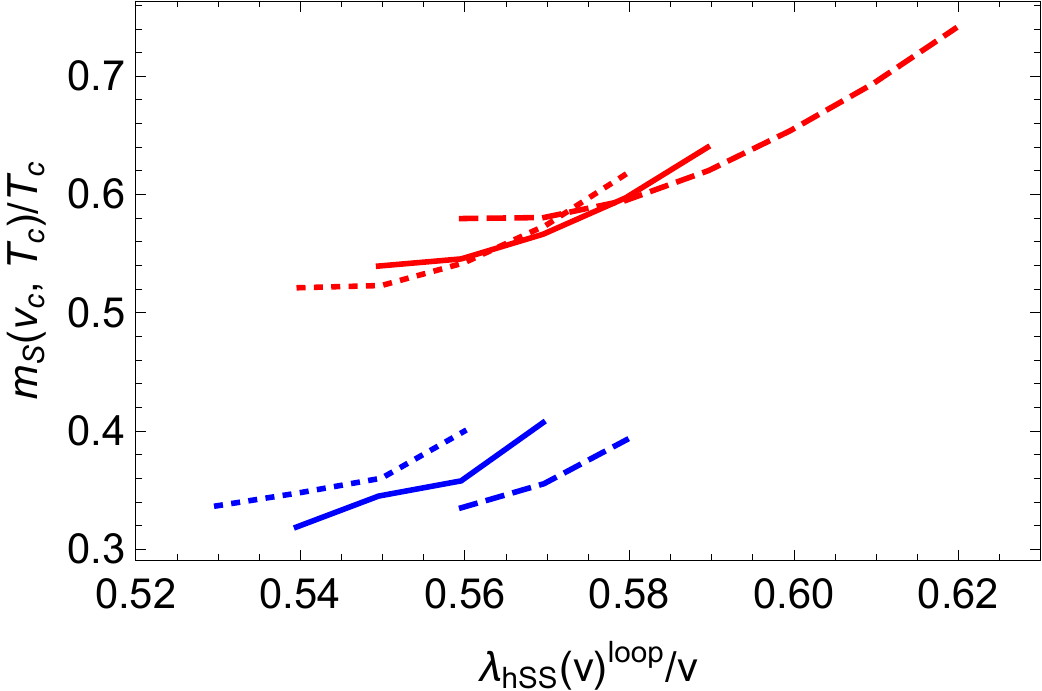}
\\
\footnotesize (b) $v_c$ && \footnotesize (f) $r_{LL}^S$&& \footnotesize (j) $m_S$ at $v_c$\\  &&&&\\
\includegraphics[width=0.3\textwidth]{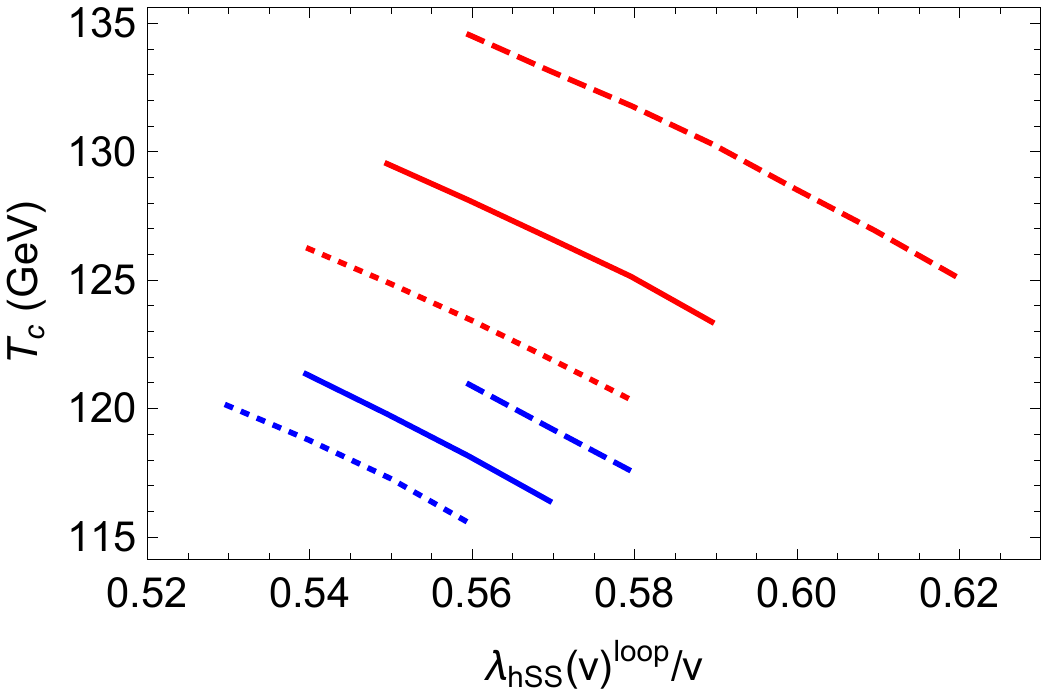}
&&
\includegraphics[width=0.3\textwidth]{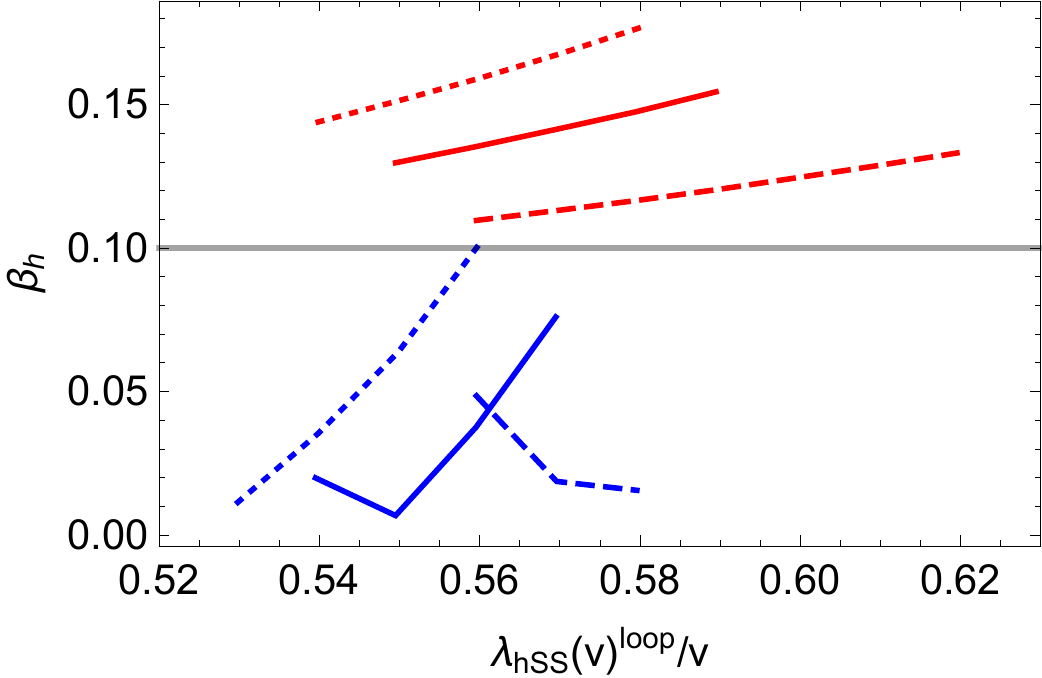}
& & 
\includegraphics[width=0.3\textwidth]{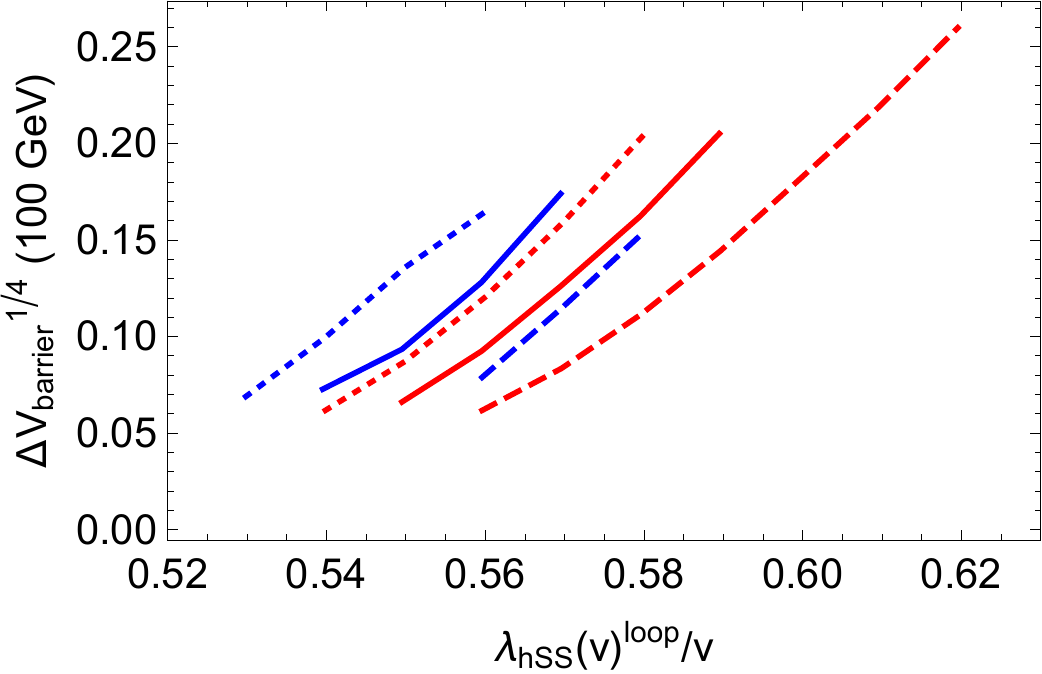}
\\
\footnotesize (c) $T_c$ && \footnotesize (g) $\beta_h$ && \footnotesize (k) potential barrier\\  &&&&\\
\includegraphics[width=0.3\textwidth]{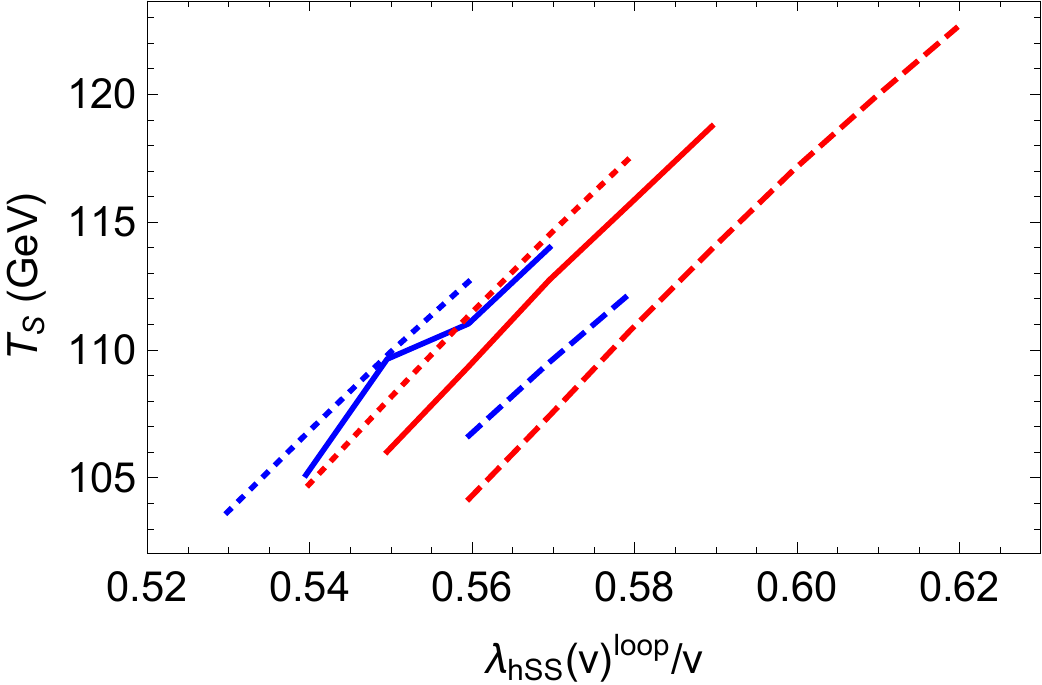}
&&
\includegraphics[width=0.3\textwidth]{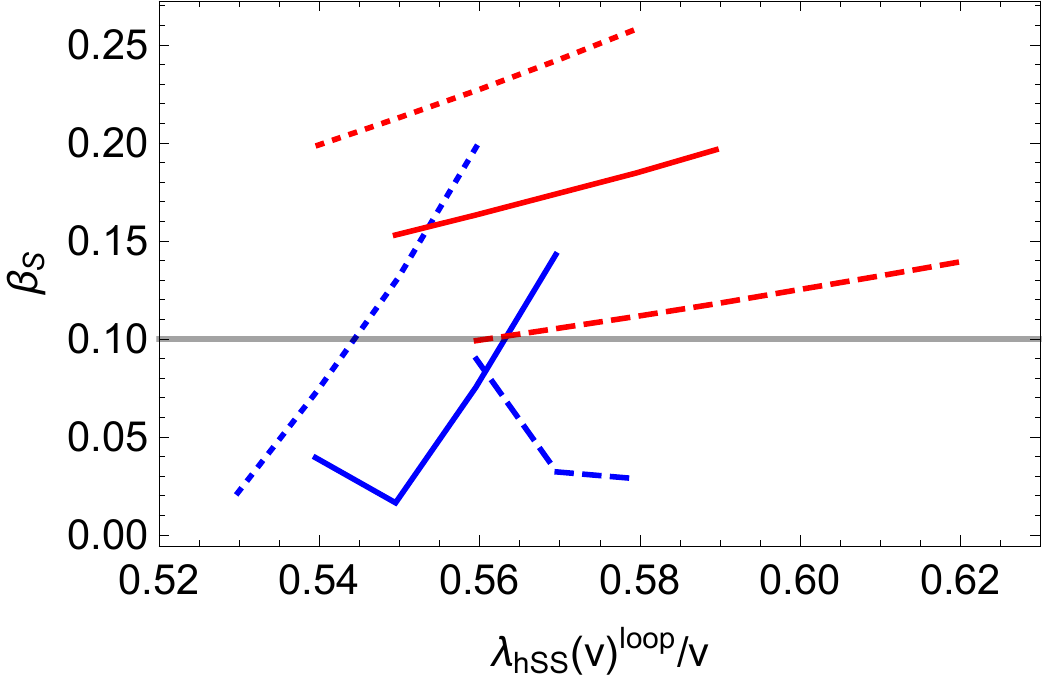}
& & 
\includegraphics[width=0.3\textwidth]{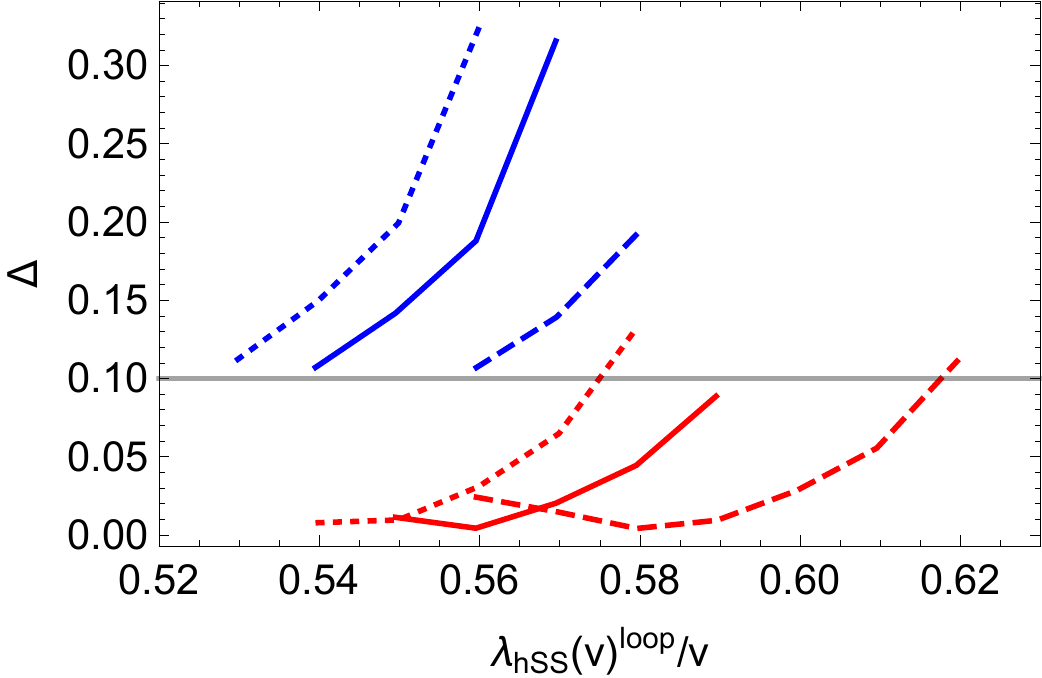}
\\
\footnotesize (d) $T_S$ && \footnotesize (h) $\beta_S$ && \footnotesize (l) Largest error term $\Delta^\mathrm{TFD, PD}$
\end{tabular}
\end{center}
\caption{
Comparison of one-step phase transition in the new PD (blue) vs the standard TFD (red) calculation, 
for $N_S = 6$ and
$(m_S, \lambda_{S}^\mathrm{loop}) = (150 \gev, 1.0)$.
The renormalization scale is set to $\mu_R = m_S$ (solid lines). Dashed (dotted) lines correspond to $mu_R = 2 m_S$ ($m_S/2$) to demonstrate the effect of scale variation. To the left of the curves, the PT is one-step and weakly first order or second order. To the right of the curves, $T_S > T_c$ and the transition is two-step for $\lambda_{hSS}(v)^\mathrm{loop} < \lambda_{hSS}^\mathrm{max}$. This upper bound is is set by the condition that EWSB vacuum is preferred and depends on $N_S, m_S, \lambda_S^\mathrm{loop}, \mu_R$ but not the choice of thermal resummation scheme.
}
\label{f.comparisonplot1D}
\end{figure}

One way to understand the different predictions of the TFD and (O)PD is to take a slice of parameter space with constant physical singlet mass $m_S$ and singlet quartic $\lambda_S^\mathrm{loop}$ in our EWSB vacuum. The strength of the phase transition, $v_c/T_c$, is shown along with several other important observables and parameters in \fref{comparisonplot1D} for $N_S = 6$ and $(m_S, \lambda_{S}^\mathrm{loop}) = (150 \gev, 1.0)$.

We first explain the qualitative features of \fref{comparisonplot1D}  which are common to both calculations. 
For very small Higgs portal coupling $\lambda_{HS} \approx \lambda_{hSS}^\mathrm{loop}/v$, the singlet sector has no effect on the EWPT, making it weakly first order or second order as in the SM. The singlet mass, which is $m_S^2 = \mu_S^2 + \lambda_{HS} v^2$ at tree-level, is given entirely by the parameter $\mu_S^2$. 
As $\lambda_{HS}$ is increased, $\mu_S^2$ decreases and eventually becomes negative to keep $m_S$ fixed. 
At some point this allows a partial cancellation between $\mu_S^2$ and $\delta m_S^2$ along the lines of \eref{cubicterm} to occur, resulting in a one-step first order phase transition starting around $\lambda_{hSS}^\mathrm{loop}/v \approx 0.55$ in \fref{comparisonplot1D} (a).
This cancellation is only partial, as shown by the plot of total finite-temperature singlet mass at the origin (i). 
Note from (c) and (d) that $T_S$, the minimum temperature at which thermal effects stabilize the singlet at the origin, is lower than $T_c$, as required for the singlet to be stable when the Higgs undergoes its one-step phase transition. 
Increasing the Higgs portal coupling drastically increases $v_c$ and hence the strength of the phase transition, see (a) and (b), since it enhances the negative ``cubic term'' of \eref{cubicterm} (from the tree-level Higgs dependence of the singlet mass) while also increasing the singlet thermal mass and therefore enhancing the cancellation of $\mu_S^2$ and $\delta m_S^2$, see (i).
For $\delta m_{h,S}^2 \sim \lambda_{HS} T^2$, both $T_c$ and $T_S$ can be schematically understood as the solution to the equations $\mu^2 = \lambda_{HS} T^2$ and $-\mu_S^2 = \lambda_{HS} T^2$ (neglecting numerical prefactors). This explains why $T_c$ decreases with increasing Higgs portal coupling, but is insufficient to understand why $T_S$ increases, since both $\mu_S^2$ and $\delta m_S^2$ depend linearly on $\lambda_{HS}$ at leading order in temperature for fixed physical singlet mass $m_S$. Solving for $T_S$ with the full high-temperature expansion of the thermal potential reproduces the behavior shown in (d). 
As the Higgs portal coupling is further increased, $T_S$ becomes larger than $T_c$, which occurs around $\lambda_{hSS}^\mathrm{loop}/v \approx 0.6$ in \fref{comparisonplot1D}. This means the phase transition is now two-step: as the universe cools it falls first into the $(h, S) = (0, w)$ vacuum before transitioning to the $(v, 0)$ vacuum. Increasing $\lambda_{HS}$ decreases the potential difference $V_\mathrm{eff}^{T=0}(v,0) -  V_\mathrm{eff}^{T=0}(0,w)$, which delays the second transition, and by lowering $T_c$ enhances $v_c/T_c$. Two-step phase transitions can therefore be very strong, since they can rely on supercooling the universe. Finally, as $\lambda_{HS}$ is increased further still the $(h, S) = (0, w)$ vacuum becomes preferred to our vacuum at zero temperature, and the model is not compatible with our universe.

With this understanding, we can now interpret the differences between the standard TFD and the new (O)PD calculation in detail:
\begin{itemize}
\item  (O)PD and TFD predict different parameter regions where the one-step phase transition is strongly first order. 
This arises due to three effects: in (O)PD, $v_c$ is larger, $T_c$ is smaller, and $T_S$ is larger than in TFD. The first and second effect make the PT stronger at a given parameter point, but the second and third effect lead to a lower Higgs portal coupling at which the switch from one-step to two-step PT occurs. 

The third effect can be traced back to the smaller thermal mass corrections obtained in (O)PD, while the first two effects are also connected to the $h$-dependence of the mass correction (see \fref{comparisonplot}). 

As a result, the region of parameter space in which (O)PD predicts a strong first order one-step PT with $v_c/T_c > 0.6$ is much smaller than in TFD, and shifted to smaller Higgs portal couplings. By the same token, the region of parameter space where the PT is two-step (to the right of the curves in \fref{comparisonplot1D}) is larger in (O)PD.

\item The finite temperature singlet mass at the origin and at $v_c$, shown in  \fref{comparisonplot1D} (i) and (j), shows that the high-temperature approximation is fairly reliable in its untruncated form (since $m/T \lesssim 1$), but the truncated high-$T$ approximation assumed in the TFD thermal mass calculation makes errors of $m/T \sim$ 30 - 70\% depending on $h$, which is consistent with \fref{comparisonplot}~(right).

\item The Higgs portal dependent curves of (b) $v_c$ and (k) potential barrier height of the (O)PD calculation are very similar to TFD curves that are shifted to lower couplings. The same holds for $v_c/T_c$, which is controlled by the rapidly-varying $v_c$. Therefore, we expect the ratio $T_n/T_c$ of the nucleation temperature, when the bubbles of true vacuum actually form, to the critical temperature to be similar in the two calculation. 

\item The absolute value of the critical temperature $T_c$ for a given strength of phase transition  $v_c/T_c$ is $\sim 10\%$ lower than predicted by the TFD calculation. If a one-step transition could be strong enough to be detected by gravitational wave observations, this would effect the frequency spectrum of the stochastic gravitational wave signal.

\item In the example of \fref{comparisonplot1D}, the overall error term (l) $\Delta^{TFD, PD}$ is dominated by the singlet lollipop ratio (f) $r_{LL}^S$. The error term is much larger in the new (O)PD calculation than the standard TFD. Since the latter underestimates the error, we expect $\Delta^\mathrm{PD}$ to give a much better representation of the calculation's reliability, which is breaking down near the switch from one-step to two-step phase transitions (larger $\lambda_{hSS}^\mathrm{loop}/v$). 

For other slices of parameter space, the $\beta$-errors can dominate, in which case the (O)PD calculation can be much more reliable, since the first neglected contributions are $\mathcal{O}(\beta^3)$. 
\end{itemize}

\begin{figure}
\begin{center}
\hspace*{-7mm}
\begin{tabular}{cc}
\includegraphics[width=0.5\textwidth]{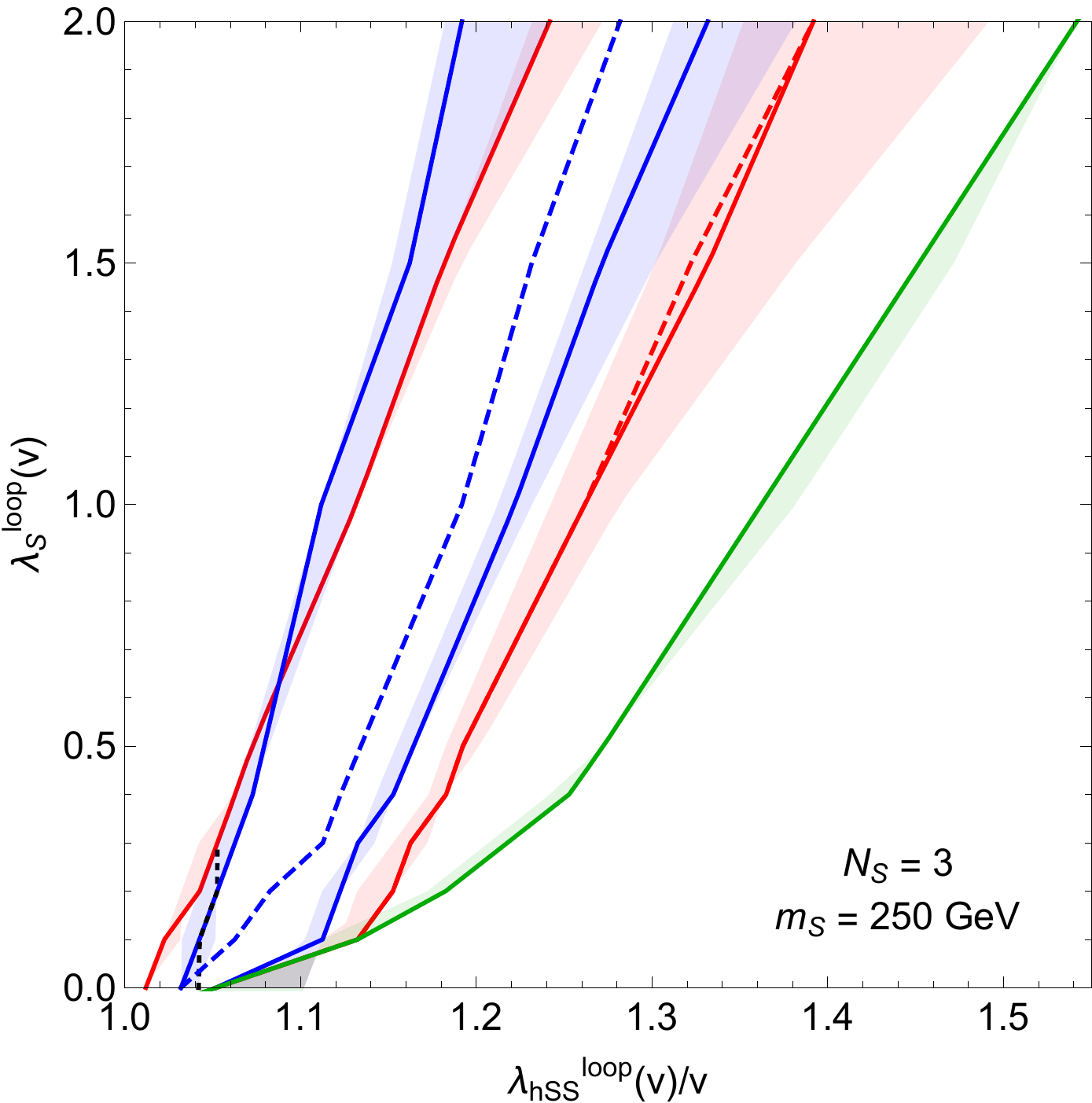}
&
\includegraphics[width=0.5\textwidth]{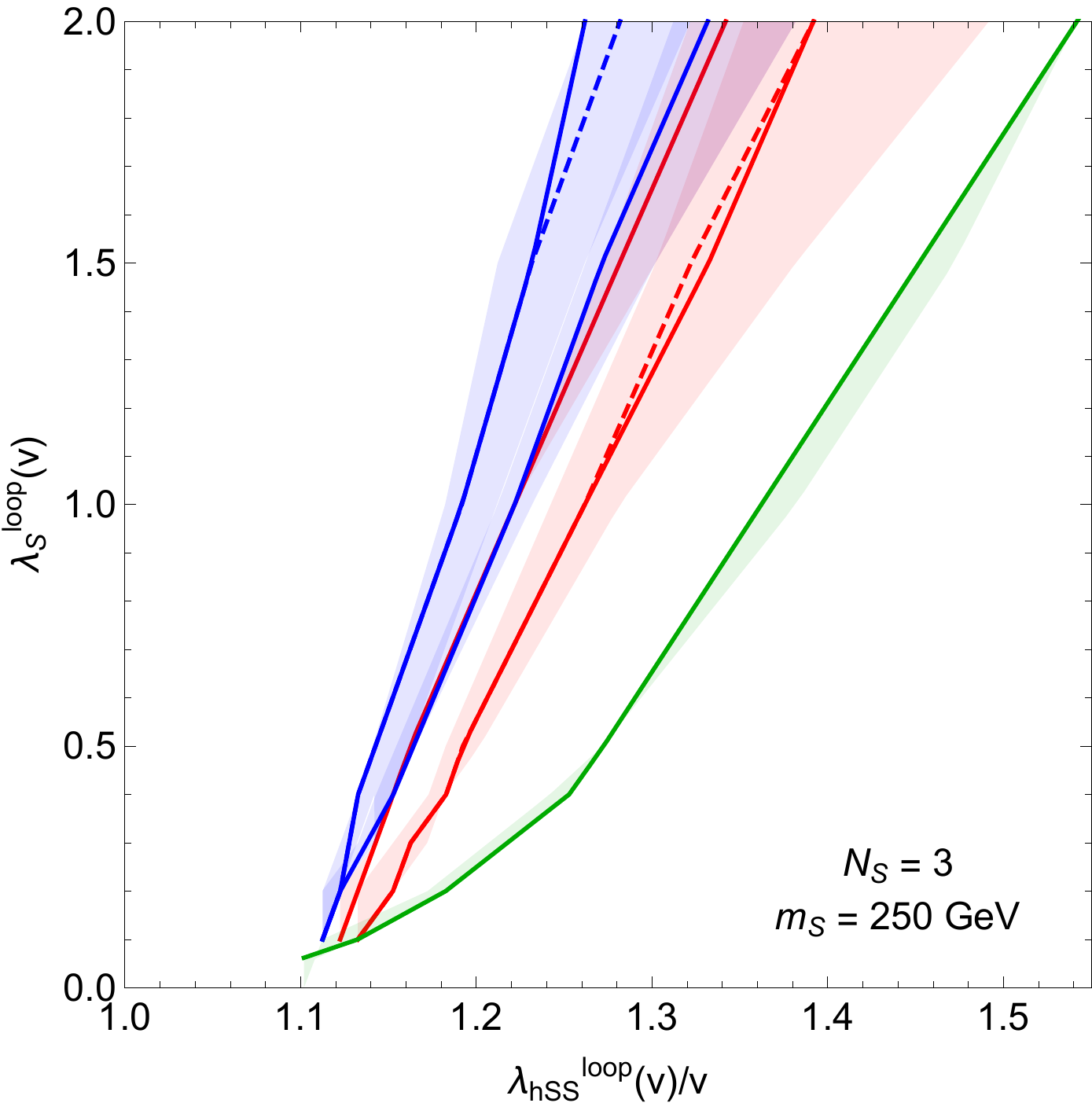}
\end{tabular}
\end{center}
\caption{SM + $N_S \times S$ parameter space with a strong EWPT for $N_S = 3$ and $m_S = 250 \gev$. 
Region between solid red (blue) lines: regions with strong one-step PT satisfying $v_c/T_c > 0.6$ for the standard TFD (new PD) calculation on the \emph{left}, and $v_c/T_c > 1.0$ on the \emph{right}.
To the left of these lines, the PT is weakly first order or second order. 
Between the red (blue) lines and the green line, the PT is two-step in TFD (PD) calculation.
 To the right of the green line, the EWSB vacuum is not preferred at zero temperature (this does not depend on the thermal mass resummation scheme). 
  Varying $\mu_R$ between 0.5 and 2 $m_S$ gives the variation indicated by the red/blue/green shading. To the right of the red (blue) dashed lines, $\Delta^\mathrm{TFD}$ ($\Delta^\mathrm{PD}$) $> 0.1$ for $\mu_R = m_S$. To the left of the black dotted line, the singlet is stable at the origin at zero temperature for $\mu_R = 1$. 
}
\label{f.comparisonplot2DlambdaHSlambdaS}
\end{figure}

It is now straightforward to interpret \fref{comparisonplot2DlambdaHSlambdaS}, which shows the regions of the $\lambda_{hSS}^\mathrm{loop} - \lambda_S^\mathrm{loop}$ coupling plane that give a strong one- or two-step first order EWPT in the TFD and (O)PD calculations for $N_S = 3$ and $m_S = 250 \gev$. 
In the left plot, a strong one-step transition is defined with the maximally permissive criterion $v_c/T_c > 0.6$. In the right plot, the criterion is slightly tightened, to the usual $v_c/T_c > 1.0$. 
In the more correct (O)PD calculation, the region allowing for a strong one-step transition (between blue lines) is smaller than in TFD (red lines), while the region with a strong two-step transition (between the blue/red lines and the green line) is larger in (O)PD. 
This is especially pronounced when $v_c/T_c$ is required to be larger than 1, in which case there is almost no overlap between the two region with a strong one-step phase transition.

\begin{figure}
\begin{center}
\begin{tabular}{cc}
\includegraphics[width=0.45\textwidth]{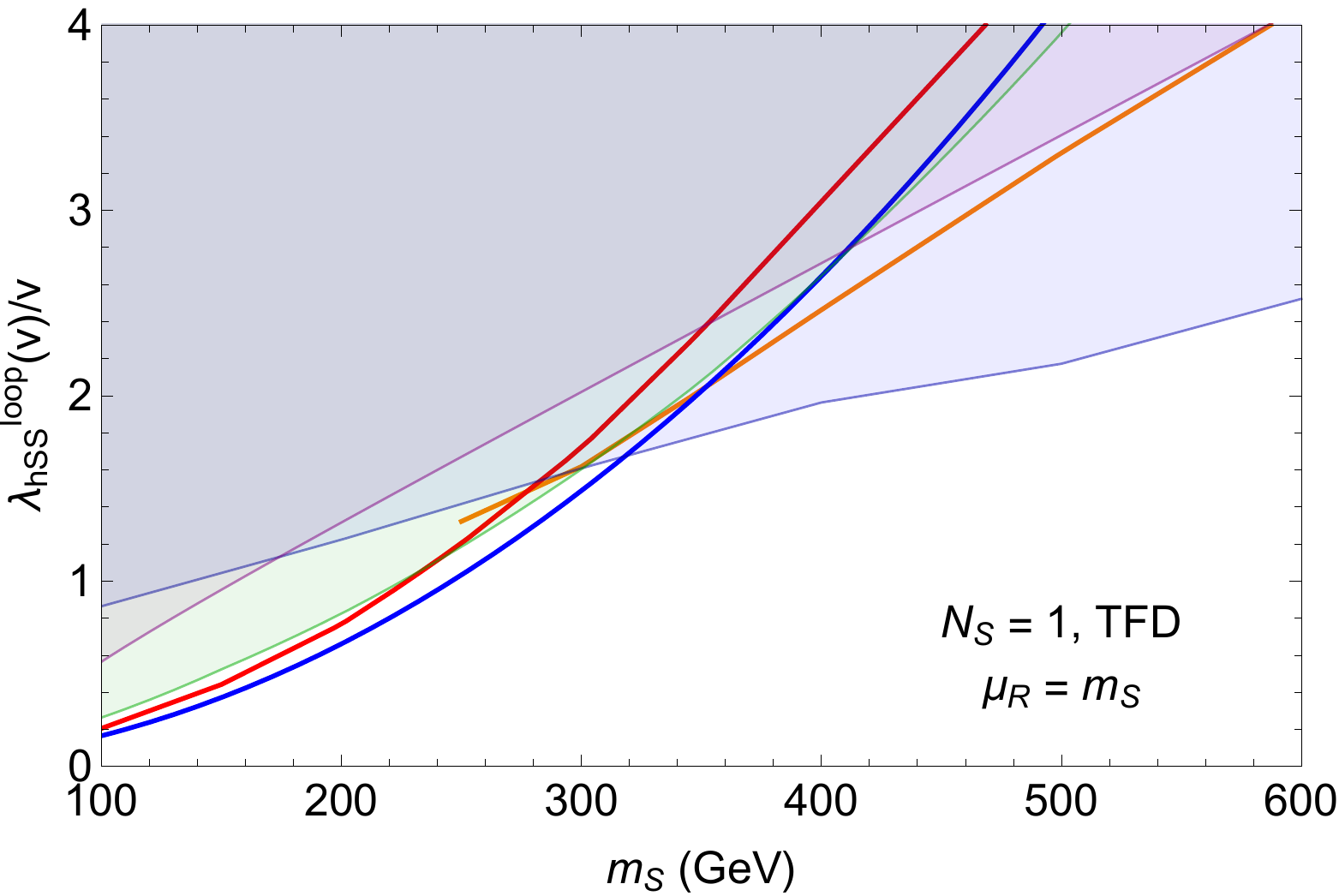}
&
\includegraphics[width=0.45\textwidth]{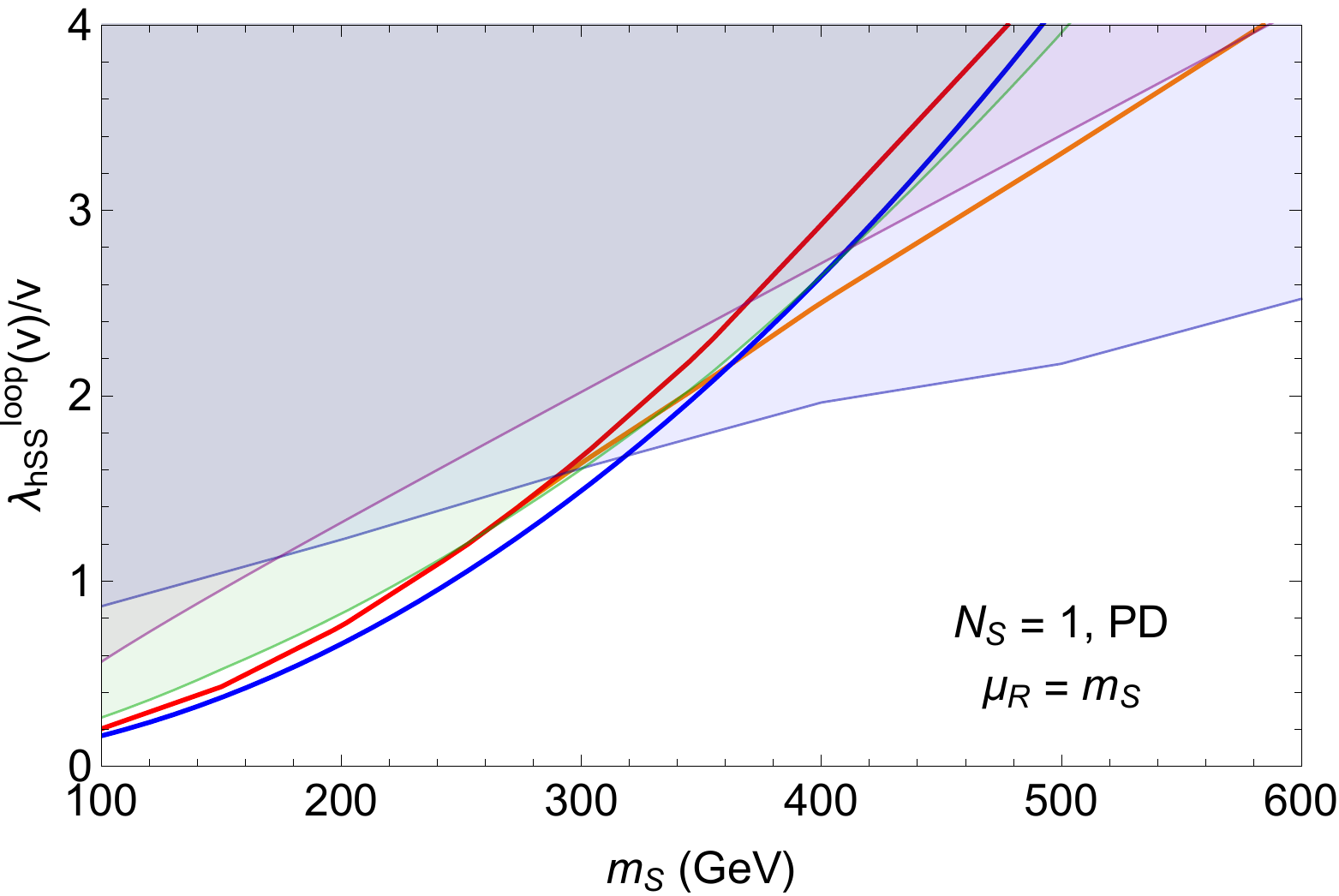}
\\
\includegraphics[width=0.45\textwidth]{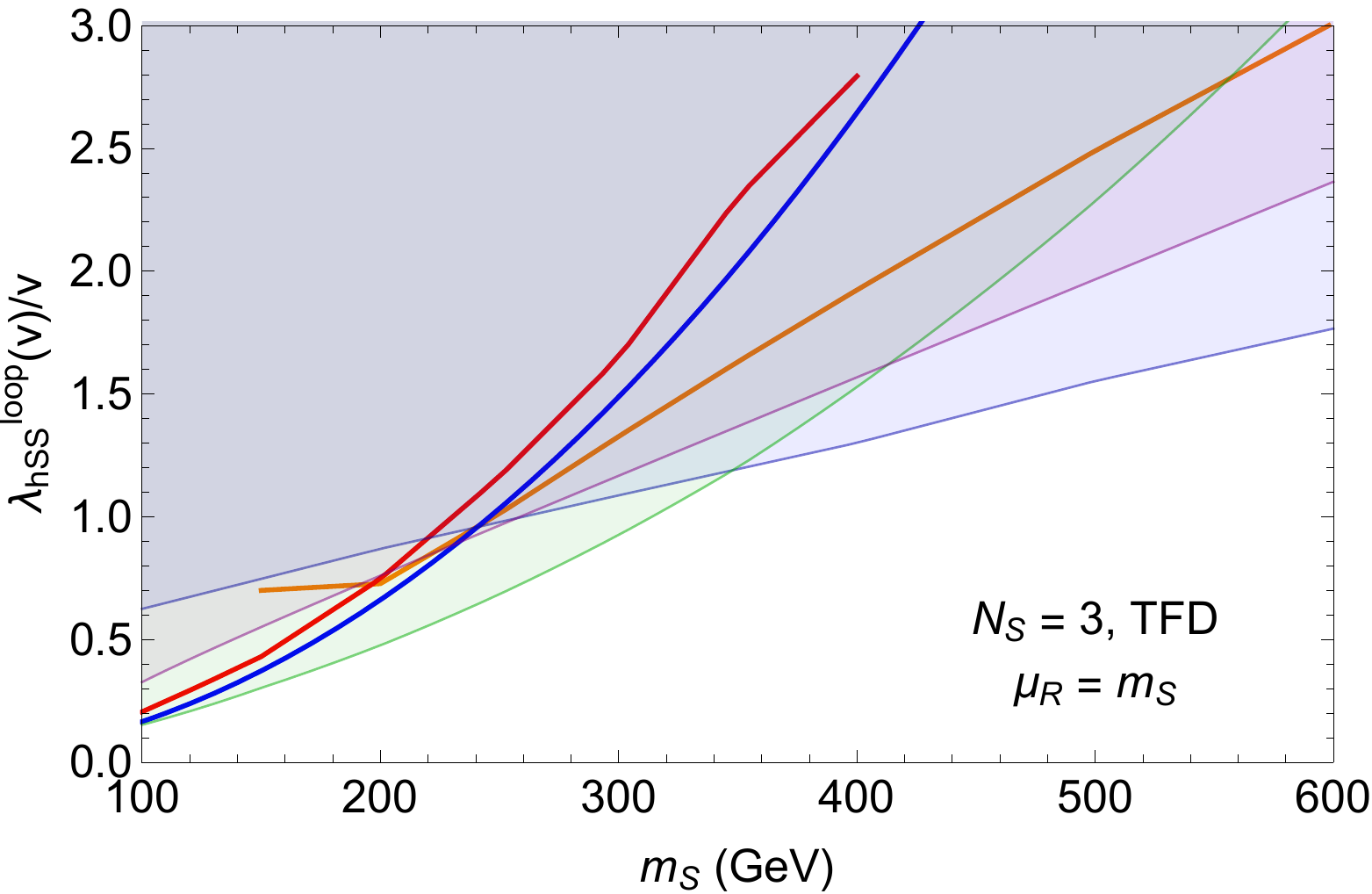}
&
\includegraphics[width=0.45\textwidth]{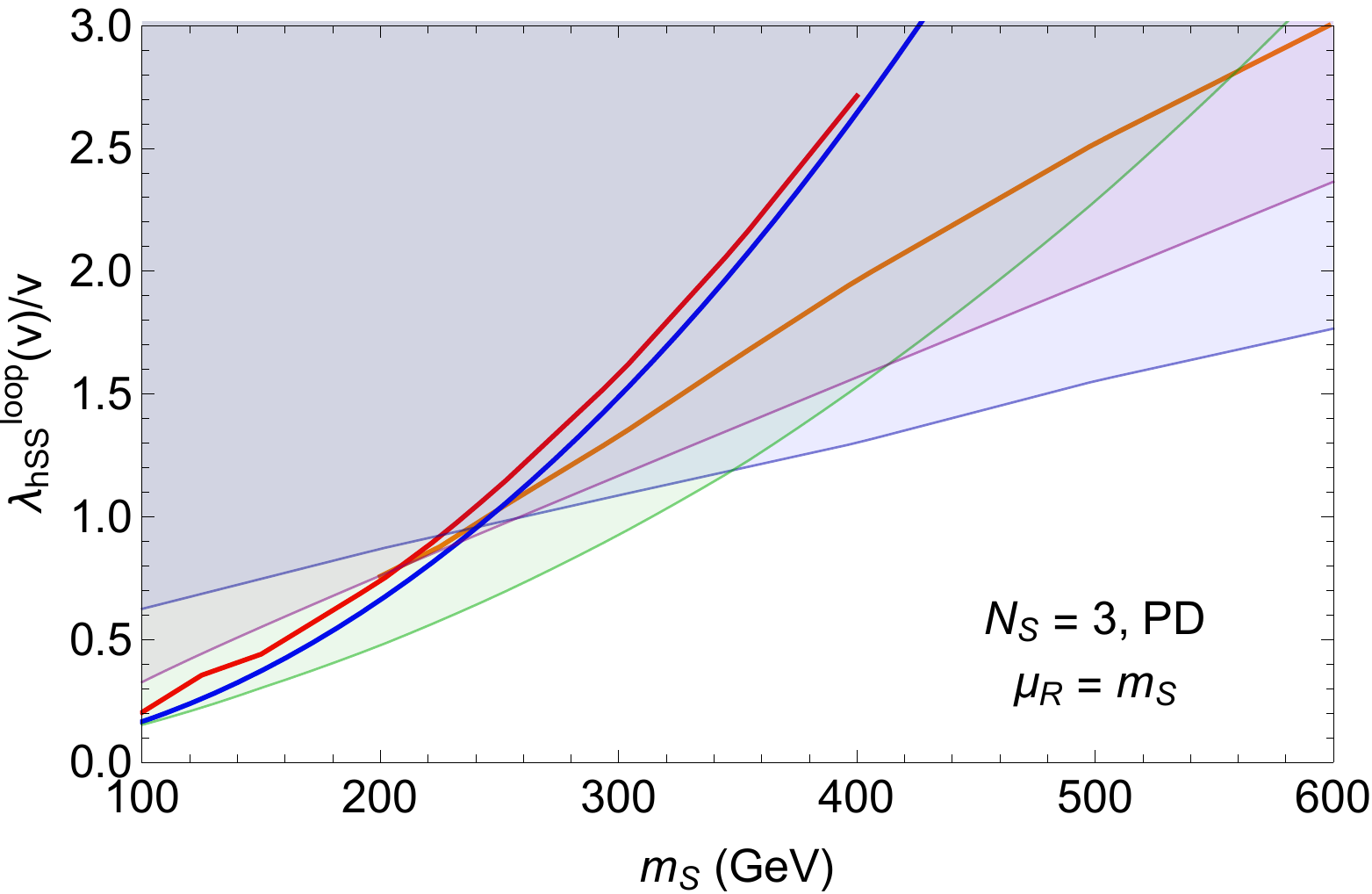}
\\
\includegraphics[width=0.45\textwidth]{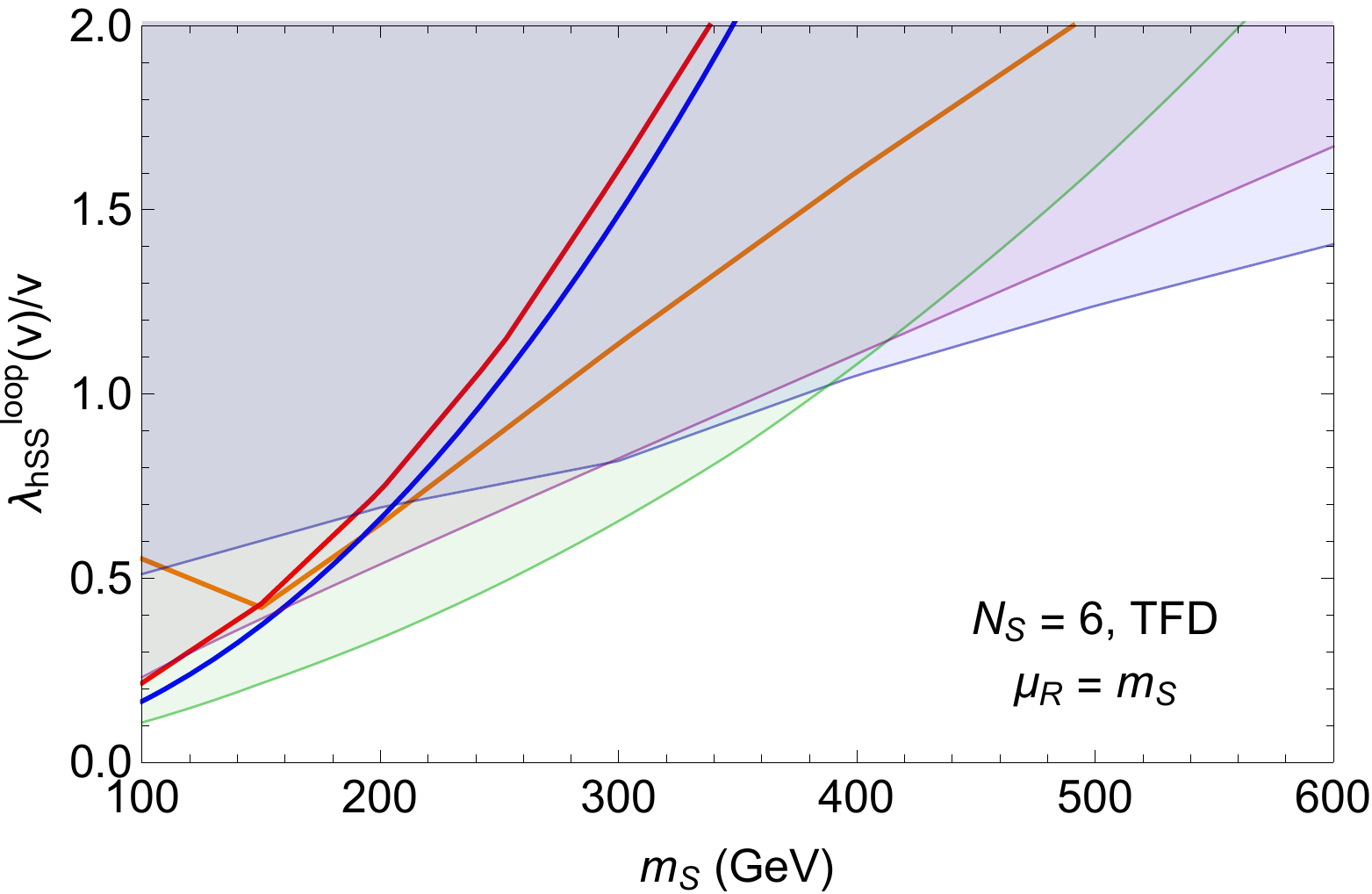}
&
\includegraphics[width=0.45\textwidth]{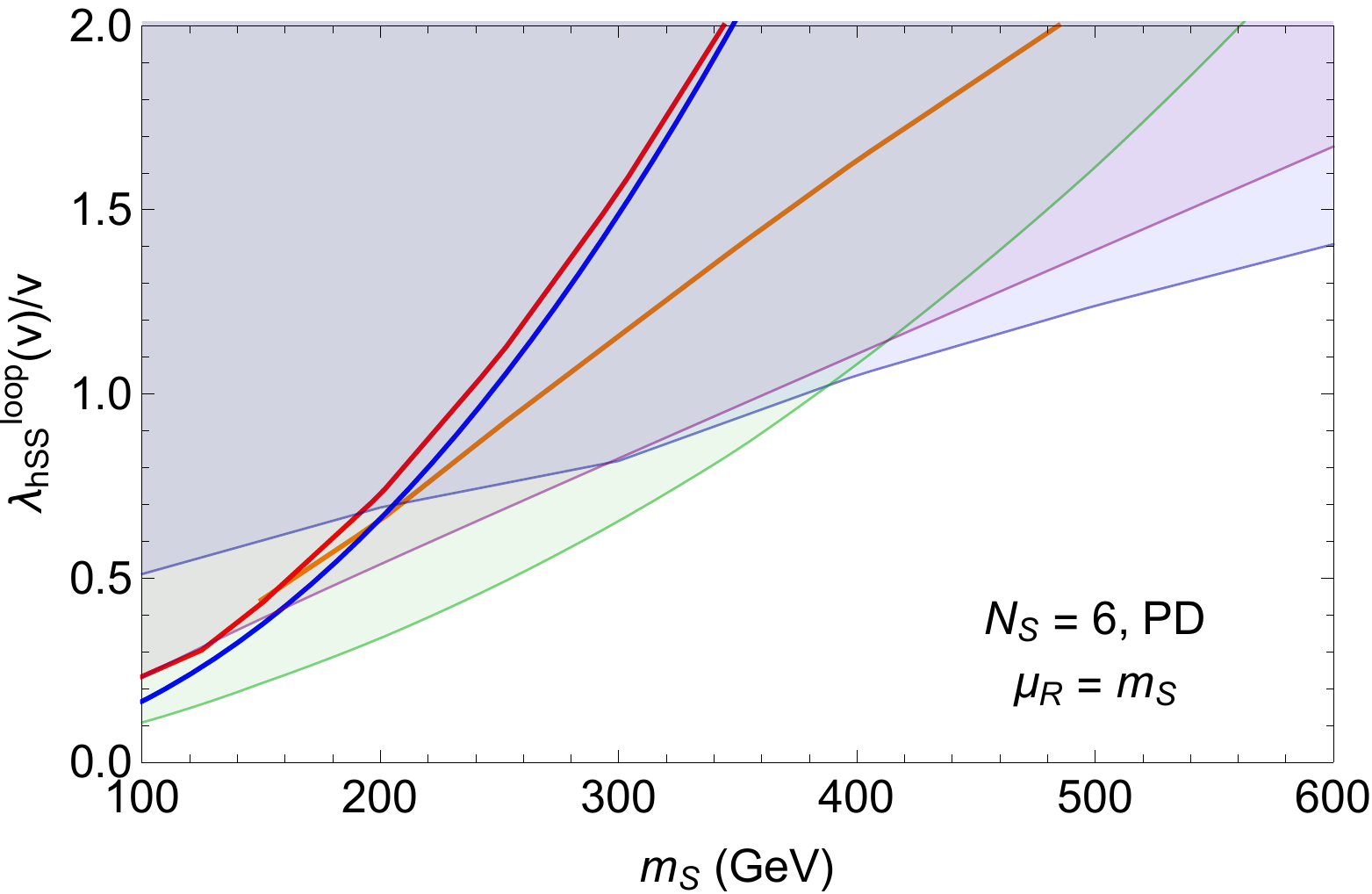}
\end{tabular}
\vspace{2mm}
\includegraphics[width=0.6\textwidth]{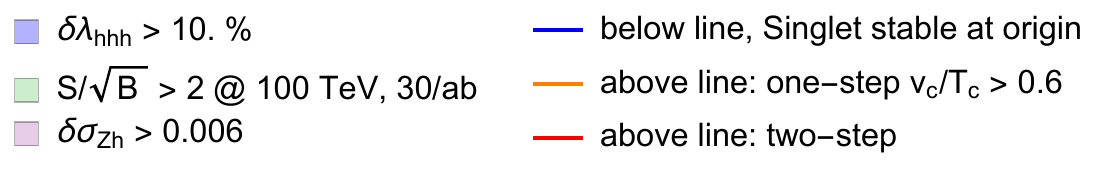}
\end{center}
\vspace*{-9mm}
\caption{
Phenomenological parameter space of the SM + $N_S \times S$ benchmark model of EWBG, for $N_S = 1, 3, 6$, computed in the TFD or PD thermal resummation scheme. The physical singlet mass in our EWSB vacuum is on the horizontal axis, the higgs-singlet cubic coupling is on the vertical axis. This determines most collider observables. 
The Higgs cubic coupling deviation \eref{lambdahhh} is bigger than 10\% in the \emph{blue shaded region}, to which a 100 TeV collider with 30 $\iab$ has more than $2 \sigma$ sensitivity~\cite{Contino:2016spe}. 
For the same luminosity, a direct VBF + MET search for invisible singlet pair production can exclude the \emph{green shaded region}~\cite{Curtin:2014jma}. 
In the \emph{purple shaded region}, a TLEP-like lepton collider can probe the $Zh$ cross section deviation, see \eref{deltaZh}, at the $2 \sigma$ level. 
Below the \emph{blue line}, the singlet is stable at the origin. 
Above the \emph{orange or red solid lines}, a one-step or two-step phase transition strong enough for EWBG can occur ($\mu_R = m_S$). 
In this projected-down parameter space, the effect of scale variation is minor. 
%%
%Dashed (dotted) red/orange lines correspond to $mu_R = 2 m_S$ ($m_S/2$) to demonstrate the effect of scale variation. 
}
\label{f.phenocomparisonplot}
\end{figure}

For the specific SM + $N_S \times S$ benchmark model, the collider phenomenology depends almost exclusively on the Higgs portal coupling and the singlet mass. \fref{phenocomparisonplot} shows the parameter regions where, for some choices of the Singlet quartic $\lambda_S$, the one- or two-step EWPT can be strong enough for EWBG. This generalizes the results of \cite{Curtin:2014jma} to $N_S \geq 1$ and makes clear that future colliders will be able to probe the entire parameter space of this representative class of models for arbitrary number of singlets.

In this particular benchmark model, the singlet quartic and the Higgs portal coupling are free parameters, and the lack of mixing with the Higgs makes the physical effects of the singlet quartic very hard to observe at colliders. Therefore, the old TFD and more correct (O)PD calculations give very similar predictions for the collider phenomenology of EWBG. However, as discussed above, the situation would be very different in more complete theories, especially if a full calculation of the baryon asymmetry reveals that $v_c/T_c$ has to be larger than 0.6. Furthermore, even in the SM + $N_S \times S$ benchmark model, there is an important difference: the minimum singlet mass required for a strong one-step PT is \emph{higher} in (O)PD than in TFD. This means that for light singlet masses, the only possibility for EWBG is via a two-step transition. As discussed in~\cite{Kozaczuk:2015owa}, these transitions can form runaway bubbles of true vacuum, which do not permit successful baryon number generation, and further study could reveal very strong additional constraints on the parameter space actually compatible with complete EWBG. Furthermore, the stochastic gravitational wave background generated by a strong two-step transition, but \emph{not} a moderately strong one-step transition, could be detected by gravitational wave observatories~\cite{Grojean:2006bp}. The (O)PD calculation reveals this exciting possibility to be more likely for light singlet masses than previously assumed.

%%%%%%%%%%%%%%%%%%%%%%%%%%%%
%%%%%%%%%%%%%%%%%%%%%%%%%%%%
%%%%%%%%%%%%%%%%%%%%%%%%%%%%
\section{Conclusions}
\label{s.conclusions}
%%%%%%%%%%%%%%%%%%%%%%%%%%%%
%%%%%%%%%%%%%%%%%%%%%%%%%%%%
%%%%%%%%%%%%%%%%%%%%%%%%%%%%

In this paper we developed the \emph{Partial Dressing} and \emph{Optimized Partial Dressing} schemes for computation and resummation of thermal masses beyond the high-temperature approximation in general BSM scenarios. This allows for the strength of  Phase Transitions to be determined to much greater accuracy than the standard Truncated Full Dressing scheme, which only resums hard thermal loops by inserting $\Pi_i \sim T^2$ into the effective potential. 

Our phenomenological analysis of the EWPT in the SM + $N_S \times S$ benchmark model generalizes the results of \cite{Curtin:2014jma}, and shows that EWBG in singlet extensions without Higgs mixing is guaranteed to be discovered at future 100 TeV and lepton colliders. Given that more general models with Higgs mixing generate additional signatures which are expected to be detectable at a future 100 TeV collider~\cite{Kotwal:2016tex}, the outlook for a general phenomenological no-lose theorem is optimistic, though more work is needed to make this conclusion completely robust.

The (O)PD calculation shows two-step phase transitions are more prevalent than previously assumed from TFD calculations. This is encouraging, as strong two-step transitions can generate observable gravity wave signals \cite{Grojean:2006bp}. They are also more constrained, since runaway bubbles are incompatible with baryon number generation \cite{Kozaczuk:2015owa}. Further analysis is needed to determine whether this translates to additional constraints on the SM + $N_S \times S$ benchmark model. 

The OPD scheme represents a simple extension on the standard TFD calculation, takes only slightly more CPU time to solve for the strength of the phase transition, and is easy to implement in Mathematica. We supply a condensed instruction manual in \aref{instructions}. We hope that the OPD calculation will be useful in the future study of other BSM scenarios. This is particularly motivated, since for more complete theories with additional correlations amongst the low-energy parameters than in our benchmark model, the OPD calculation makes significantly different predictions for the EWBG-viable parameter space, and hence the associated collider and cosmological observables. In some cases, scenarios which were thought to be viable may now be excluded.

Developing the (O)PD thermal resummation scheme is a necessary ingredient for the careful study of EFTs at higher temperature, which in turn would represent a great leap in our model-independent understanding of EWBG. We are currently conducting such an analysis, and will present the results in a future publication.

\acknowledgments
We thank
Thomas Konstandin,
Jonathan Kozaczuk,
David Morrissey,
Jose Miguel No,
Hiren Patel,
Michael Ramsey-Musolf,
Carlos Tamarit
and the participants of the MIAPP workshop ``Why is there more Matter than Antimatter in the Universe?'' 
for useful discussions. 
The work of DC was supported in part by the Munich Institute for Astro- and Particle Physics (MIAPP) of the DFG cluster of excellence "Origin and Structure of the Universe".
D.C. is supported by National Science Foundation grant No. PHY-1315155 and the Maryland Center for Fundamental Physics.
P.M and H.R. are supported in part by NSF CAREER award NSF-PHY-1056833 and NSF award NSF-PHY-1620628.

\appendix

%%%%%%%%%%%%%%%%%%%%%%%%%%%%
%%%%%%%%%%%%%%%%%%%%%%%%%%%%
%%%%%%%%%%%%%%%%%%%%%%%%%%%%
\section{Instruction Manual for Optimized Partial Dressing Calculation of Phase Transition}
\label{a.instructions}
%%%%%%%%%%%%%%%%%%%%%%%%%%%%
%%%%%%%%%%%%%%%%%%%%%%%%%%%%
%%%%%%%%%%%%%%%%%%%%%%%%%%%%

Here we summarize the detailed procedure for obtaining the effective finite-temperature potential $V_\mathrm{eff}^{\mathrm{pd}}(h,T) $ as a function of $h$ for a given temperature $T$ in the Optimized Partial Dressing (OPD) scheme. 
 This will be familiar to anyone studying the EWPT in BSM theories, and implementing the OPD calculation in Mathematica is very similar to the familiar TFD calculation, and only $\sim \mathcal{O}(10\%)$ more CPU intensive. 
 We explain this procedure in the context of the SM + $N_S \times S$ benchmark model, but it generalizes easily to other theories with one-step phase transitions. At every point, only use the real parts of various potential contributions or their derivatives. Note that we do not perform this resummation for zero temperature matching and other calculations, since the effect is small. 
 
For a given temperature, we first have to find the thermal mass corrections $\delta m_i^2(h,T)$. For gauge bosons, use the standard $\delta m_i^2 = \Pi_i \sim T^2$, see e.g. \eref{PiT2} for the SM contributions. For scalars we have to solve a set of coupled gap equations:
\begin{equation}
\label{e.BSMgapeqn2app}
\delta m_{\phi_j}^2(h, T) =
\sum_i
\frac{\partial}{\partial \phi_j}
 \left[ 
\frac{\partial V^i_\mathrm{CW}}{\partial \phi_j}\Big(m_i^2(h) + \delta m^2_i(h, T)\Big) + \frac{\partial V^i_\mathrm{th}}{\partial \phi_j}\Big(m_i^2(h) + \delta m^2_i(h, T),T\Big)
\right]
\end{equation}
See Eqns.~(\ref{e.CW0}) and ~(\ref{e.Vth}) for definitions of the Coleman-Weinberg and finite-temperature potential. For the latter, use the high-temperature expansion to log-order in the gap equation, see \eref{JBFhighT}. Note that this makes the RHS a set of fully analytical expressions that can be easily manipulated in Mathematica. 

In our specific benchmark model, there are three gap equations: for $j = h, G_0, S_0$. Since we are interested in excursions along the $h$-axis, all Goldstone thermal masses are the same but different from the physical Higgs mass, and all scalar thermal masses are the same.\footnote{If we solved for excursions along the $S_0$ direction, we would have to treat $S_0$ and $S_{k>0}$ differently but could treat all Higgs degrees of freedom the same.} Note that the sum $i$ also runs over all particles, including fermions and gauge bosons. 

We numerically solve this set of gap equations by essentially treating it as a set of partial differential equations. Specifically, set up a coarse grid\footnote{To avoid singularities the origin can be defined to be a very small positive displacement instead of identically zero.} along the Higgs axis $h = \{0, 10, 20, \ldots, 250\}$. 
The RHS of the gap equation is a function of thermal masses and their derivatives.  Expanding the mass correction around $h = h_a$:
\begin{equation}
\delta m_j^2(h,T) \approx \delta  m_{j(a)}^{2} + (h - h_a) \frac{\partial \delta  m_{j(a)}^{2}}{\partial h}
\end{equation}
and substituting this form of $\delta m_{h,G_0,S}^2$ (as well as the $h$-independent gauge boson thermal masses) into the RHS yields three gap equations that depend on six parameters: the value of the scalar thermal masses $\delta  m_{h(a)}^{2}, \delta  m_{G_0(a)}^{2}, \delta  m_{S(a)}^{2}$  and their first derivatives $\partial \delta  m_{h(a)}^{2}/\partial h, \partial \delta  m_{G_0(a)}^{2}/\partial h, \partial \delta  m_{S(a)}^{2}/\partial h$.

At the origin, the derivatives are zero, and solving for just the three thermal masses yields a numerically unique set of solutions. Then we work our way away from the origin. The solution at $h = h_a$ can be obtained by first eliminating either the derivatives or the actual thermal masses from the gap equation using the continuity condition
\begin{equation}
\label{e.continuityconditionapp}
\delta  m_{j(a-1)}^{2} \approx \delta  m_{j(a)}^{2} + (h_{a-1} - h_{a}) \frac{\partial \delta  m_{j(a)}^{2}}{\partial h}
\end{equation}
given the known solution at $h = h_{a-1}$. We found in practice that eliminating the $ \delta  m_{j(a)}^{2} $ and solving for the three derivatives ${\partial \delta  m_{j(a)}^{2}}/{\partial h}$ was more reliable. If the solution fails the resolution of the grid may have to be increased, but we found a coarse grid with 10 GeV spacing to be sufficient.\footnote{The automatic Mathematica functions for differential equation solving failed with these gap equations, possibly by over-interpreting unphysical local singularities or numerical noise. We found our manual implementation to be very simple, fast and reliable.}

Once a grid of solutions is obtained, the continuous functions $\delta m_j^2(h,T) $ are defined by linear interpolation. The thermal potential is then defined by
\begin{equation}
\label{e.VefffiniteTdressedapp}
V_\mathrm{eff}^{\mathrm{pd}}(h,T) = V_0 + 
 \sum_i 
 \int dh
 \left[ 
\frac{\partial V^i_\mathrm{CW}}{\partial h}\Big(m_i^2(h) + \delta m^2_i(h, T)\Big) + \frac{\partial V^i_\mathrm{th}}{\partial h}\Big(m_i^2(h) + \delta m^2_i(h, T),T\Big)
\right]
\ ,
\end{equation}
where $i$ runs over all scalars, gauge bosons and fermions (with zero thermal mass for the latter). Importantly, in this effective potential, use the piece-wise defined thermal functions
\begin{eqnarray}
\nonumber
J_B^\mathrm{piece-wise}(y^2) &=& \left\{
\begin{array}{ll}
J_B^{\mathrm{high}-T}(y^2)  & \mathrm{for} \ y^2 \leq 1.22
\\
\tilde J_B^{(3)}(y^2) & \mathrm{for} \ y^2 > 1.22
\end{array}
\right.
\\
\label{e.JBFpiecewiseapp}
\\
\nonumber
J_F^\mathrm{piece-wise}(y^2) &=& \left\{
\begin{array}{ll}
J_F^{\mathrm{high}-T}(y^2)  & \mathrm{for} \ y^2 \leq 1.29
\\
\tilde J_F^{(2)}(y^2) & \mathrm{for} \ y^2 > 1.29
\end{array}
\right.
\end{eqnarray}
See Eqns.~(\ref{e.JBFhighT}) and (\ref{e.JBFlowT}) for the definition of the approximate thermal functions. The integrand can then be defined as an analytical function in Mathematica, which can be evaluated on a grid of Higgs values $h = \{0, 5, 10, \ldots, 250 \gev\}$ and interpolated to efficiently perform the integral at arbitrary values of $h <250 \gev$.

\bibliographystyle{unsrt}
\bibliography{bibliography}{}

\end{document}